\documentclass[draft,12pt]{article}
\usepackage{amsmath,amsfonts,palatino,amsthm,epsf}
\newcommand{\beq}{\begin{equation}}
\newcommand{\eeq}{\end{equation}}

\newcommand{\f}{\begin{equation}}
\newcommand{\ff}{\end{equation}}
\newcommand{\blankline}{\vskip .3cm}

\begin{document}

\title{An invitation to  Loop Quantum Gravity}
\author{ Lee Smolin\thanks{Email address:lsmolin@perimeterinstitute.ca}
\\
\\
Perimeter Institute for Theoretical Physics,\\
35 King Street North, Waterloo, Ontario N2J 2W9, Canada, and \\
Department of Physics, University of Waterloo,\\
Waterloo, Ontario N2L 3G1, Canada\\}
\date{\today}
\maketitle

\begin{abstract}

We describe the basic assumptions and key results   
of loop quantum  gravity, which is a background independent  approach to quantum gravity. 
The emphasis is on the basic physical
principles and how one deduces predictions from them, at a level
suitable for physicists in other areas such as string theory,
cosmology, particle physics, astrophysics and condensed matter physics. 
No details are given, but references are provided to guide 
the interested reader to the literature.  The present state of
knowledge is summarized in a list of  $42$ key results on topics including
the hamiltonian and path integral quantizations, coupling to matter, 
approaches to unification, 
 extensions to supergravity
and higher dimensional theories, as well as applications to
black holes, cosmology and Plank scale phenomenology. 
We describe  the near term 
prospects for observational tests of quantum theories of gravity and the 
expectations that loop quantum gravity may provide predictions for 
their outcomes. Finally, we  provide  answers to frequently asked
questions and a list of key open problems.

\end{abstract}
\vskip 1 in
{\it To be submitted to Reviews of Modern Physics}
 \newpage
\tableofcontents
\newpage

\section{Introduction}
  
Loop quantum gravity is a conservative  approach to quantum gravity that asks a  
simple question: {\it Can we construct a quantum theory of 
spacetime based only on the experimentally well confirmed principles 
of general relativity and quantum mechanics?}  Most approaches to quantum gravity
have concluded or assumed that this could not be done.  However, in the mid 1980's the
situation changed because it was understood that general relativity could be most simply formulated
as a kind of gauge theory\cite{sen,abhay}.  Given the tremendous advances in our
understanding of gauge theories that had occurred in the decade previous, this made it
possible to take a fresh approach to quantum gravity.   
  
  Remarkably, after almost two decades of work by a community of more than a  hundred physicists and mathematicians, a great deal of evidence has 
  accumulated that the answer to the question above is yes.  
The result  is a language and
dynamical framework for studying the physics of quantum spacetimes, which is completely
consistent with the principles of both general relativity and quantum field theory.  
The picture of quantum spacetime geometry which emerges  is to many compelling, independently of the 
  fact that it has been derived from a rigorous quantization of 
  general relativity. The basic structure that emerges is of a new class of quantum
gauge field theories, which are background independent, in that no fixed spacetime metric 
is needed to describe their quantum dynamics\cite{abhaybooks}-\cite{alejandro-review}. 

While a number of  issues remain open, enough is known about the physics
of loop quantum gravity that a number of applications to problems of physical
interest are under development. There is by now a well understood detailed description of the
quantum physics of 
black hole and cosmological horizons\cite{linking,kirill1,isolated} 
that reproduces the Bekenstein-Hawking
results on the relationship between area and entropy\cite{bb}.  There is under 
development an approach to quantum cosmology that shows 
that cosmological singularities are eliminated\cite{LQC}
and has already 
led to some predictions for effects observable in CMB
spectra\cite{LQC,lqc-inflation}. And there are published predictions 
for observable Planck scale deviations from
energy momentum relations\cite{GP,AMU}  
that imply predictions for experiments in progress such as
AUGER and GLAST.  For those whose interest is 
towards  speculations
concerning supersymmetry and higher dimensions, 
there are also results that show how the methods of loop 
quantum gravity may be extended to give background 
independent descriptions of quantum gravity in the higher and super realms\cite{yime-holo}-\cite{Mlee}.  
It thus seems like a good time
for an introduction to the whole approach that may help to make the basic ideas, results
and methods accessible to
a wider range of physicists. 
 
The main purpose of 
  this review is to explain how it has been possible to obtain exact 
  results about quantum gravity, describe the main features of the resulting
theory, and then to 
  list its basic results.  The text is aimed at physicists who are not specialists in quantum gravity, particularly those in elementary particle physics, string theory,
cosmology, high energy astrophysics and  condensed matter physics.  
I assume only that the reader if familiar 
 with the basics of gauge fields, quantum theory and  general relativity.  

Like any mature field, work in loop quantum gravity proceeds through a variety of methods and 
levels of rigor.  The first results were found using direct extensions of methods from ordinary gauge theories, with modifications in regularization methods appropriate to theories without
background metrics.  More recently, all the key results have been confirmed using 
mathematically rigorous methods\cite{thomas-thesis,abhayjerzy}.  In this review I employ a non-rigorous style, grounded on intuitions about the behavior of gauge fields that particle physicists and condensed matter physicists will find familiar. But the reader should keep in mind that,  as will be noted, many of  the key results listed are confirmed by completely rigorous methods.  This is good, as we have so far no experimental checks on the behavior of gauge fields in this context, it is essential to be able to confirm our intuitions with rigorous results.  

In the next section  we describe the four basic observations which lead to loop quantum
gravity.  This is followed by an intuitive description of the basic physical picture. 
In section 4 we give a list of 42 key results achieved so far concerning the theory. 
 We then 
  summarize the near term experimental situation for loop quantum 
  gravity and other theories in section 5. Section 6 gives answers to a list of
 questions most frequently asked by those curious or skeptical about loop quantum gravity, 
after which we  close with a list of open problems.   
  
  For those interested in a more detailed introduction to loop quantum 
  gravity, there are 
  several textbooks and  monographs  
available\footnote{Early 
  monographs are \cite{abhaybooks,PG-book}. A recent textbook by 
Rovelli\cite{Carlo-book}  is from Cambridge University Press, with an early draft available on 
  line at 
http://www.cpt.univ-mrs.fr/~rovelli/rovelli.html. 
  Comprehensive introductions to the rigorous side of the subject are 
  given in \cite{thomas-thesis}, also soon to appear from CUP, as well as the very recent
review\cite{abhayjerzy}.}   Review papers\footnote{References here are not intended to be complete,
and are meant only as an introduction to some key results in the 
literature.  Reminders of references I have forgotten will be appreciated.}  on 
  different aspects of loop quantum gravity  include \cite{abhayjerzy}-\cite{alejandro-review}.

\section{The four basic observations} 
  
  While the principles assumed are only those of general relativity 
  and quantum mechanics, there are four key observations that make 
  the success of loop quantum gravity possible.  These are
  
   \blankline
  
{\bf I.} {\it  Classical general relativity is a background 
      independent theory,  hence any theory which is to have general 
      relativity as a low energy limit must be background 
      independent.} A background independent theory is one whose 
      formulation does not assume 
      or require the existence of any single preferred spacetime 
      metric or connection. Instead, all the fields that define the 
      geometry of spacetime are fully dynamical, none are fixed\footnote{The reader 
should be aware that the word ``background independence" has other connotations 
besides that given here. More about this as an answer to a FAQ.}. 

The argument that a quantum theory of gravity must be background independent
can be found developed in several places\cite{Carlo-book,background}, where
one can also find the term carefully defined.  

The principle of background independence 
does not imply that the fundamental theory must be based on fields living on a manifold.
But it does imply that whenever manifolds appear, whether fundamentally or as part of
the low energy effective theory, the dynamics of the fields will be invariant under 
active diffeomorphisms
of the manifold.  The reason for this was worked out first in detail by Dirac,
and his \cite{Dirac-book} is still a good source for understanding 
the connection between background independence and 
diffeomorphism invariance\footnote{This is also described in detail in 
\cite{Carlo-book} and in \cite{stachel}-\cite{LOTC}.}.

When we say that LQG realizes the basic principles of general relativity,
we are referring to only the big principles, which include background
independence, diffeomorphism invariance and the equivalence principle.  While many papers in LQG are concerned with the quantization
of the Einstein action, we can equally well study other actions, including supergravity and 
terms of higher powers
in curvature.  Loop quantum gravity is
perfectly compatible with the expectation that the Einstein equations are just the low
energy limit of a more fundamental theory. 
      
      \blankline
      
      {\bf II.} {\it  Duality and 
      diffeomorphism 
      invariance may be consistently combined in a quantum theory.}  By duality we mean here
the conjecture that the dynamics of a quantum gauge field can be described 
equivalently in terms of the dynamics of one dimensional extended objects.  
In the context of Yang-Mills theory this conjecture was explored in detail by 
Polyakov, Migdal, Mandelstam, Nielson and others, where the one 
dimensional objects were Wilson loops.  More recently, in the background
dependent context, the idea of duality between gauge fields and extended objects has been 
central string theory, for example in the $AdS/CFT $ correspondence.   In the background independent context,
the idea is central to loop quantum gravity. As we will see, the key point is a method
for constructing a diffeomorphism invariant quantization of any  diffeomorphism invariant 
      gauge field theory.  This cannot be done using Fock states, as the inner product on Fock space depends on a background metric, whose presence breaks diffeomorphism invariance. But, as
I will describe, it can be done if one works in a space of states created by the
action of Wilson loops.  

      Indeed, from a technical point of view, the main achievement of loop 
      quantum gravity is the discovery of a new kind of quantum gauge 
      field theory, which is exactly invariant under diffeomorphisms 
      of a manifold.   This is understood in a great deal of detail, 
      and the Hilbert space, inner product, states, observables and 
      path integral representation
      are all understood in closed form. That understanding is 
      robust, and has been achieved by several regularization 
      procedures. Significantly, there is a completely rigorous formulation 
      of these diffeomorphism invariant gauge theories. A recent uniqueness
theorem\cite{unique} greatly limits the possibilities for quantum descriptions of diffeomorphism
invariant gauge theories apart from the one studied in LQG.  
      
       \blankline
      
      {\bf III.} {\it General relativity and all related 
      theories, such as supergravity, can be formulated as gauge 
      theories.} This means that the configuration variable is a 
      gauge field, and the metric information is contained in the 
      conjugate momenta to the gauge field. This extends to coupling 
      with all the known kinds of matter fields. 
      
       \blankline
      
      {\bf IV.} {\it Further, general relativity, supergravity
      and related theories can be put in a special form in which they 
      are} {\it \bf{ constrained topological field theories.}}  These are 
	defined in section 3.3, briefly, they are 
      theories whose actions differ from the action of a topological 
      field theory\footnote{A topological field theory is a field theory whose
equations of motion are all trivial, so that there are no local degrees of freedom. The solutions
are parameterized instead entirely by topological and boundary information.}
by the imposition of a non-derivative, quadratic 
      constraint equation. That constraint diminishes the number of 
      gauge invariances, leading to the emergence of local physical degrees 
      of freedom, while maintaining the diffeomorphism invariance of 
      the theory.   This is true of general relativity in all 
      dimensions and is also known to be true of supergravity, in 
      $d=4$, at least up to $N=2$\cite{yime-holo}-\cite{superyi}, 
and for $d=11$ \cite{11d}.  
      
       \blankline

      As a result, the techniques 
      which give us consistent diffeomorphism invariant gauge 
      theories can be applied to give consistent quantizations of 
      general relativity and supergravity. Many results follow, which will be 
      described in more detail below. Most of the key results, 
     have been confirmed as  
      theorems in the rigorous formulation of diffeomorphism invariant 
      quantum field theory.

     The fundamental result which follows from these four observations is
     that {\it quantum geometry is discrete}\footnote{Precise statements and
references for all these results are given in section 4.}.
      Operators which measure the areas and 
      volumes of diffeomorphism invariantly defined surfaces and 
      regions may be constructed. They are finite, after an 
      appropriate regularization, respecting diffeomorphism 
      invariance, and they have discrete, computable spectra. Hence, 
      the theory predicts minimal physical areas and volumes. By 
diagonalizing 
these observables  we find an orthonormal 
      basis of diffeomorphism invariant states, which are certain 
      labeled graphs called spin networks.  Their evolution can be 
      described in closed form in either a Hamiltonian or path 
      integral language.  Furthermore, the dynamics of the matter 
      fields can be constructed and studied. Because of the existence 
      of minimal quanta of volume, all divergences of ordinary quantum 
      field theories are eliminated. This is because there simply are 
      no degrees of freedom in the exact theory that correspond to 
      gravitons or other quanta with wavelength shorter than the 
      Planck length.   There is evidence as well that 
      singularities of general relativity are eliminated.

 It should be apparent from this summary that loop quantum gravity, 
 as a research program, is
  rather different from other research programs such as string 
  theory, that are based on new hypotheses about nature, such as 
  supersymmetry, the existence of higher dimensions, and the 
  unification of all elementary particles and forces by means of 
  strings.  These are interesting 
  hypotheses, but it is proper to characterize them as speculative, 
  as they have no direct support from experiment. Nor are there results that 
  show that any of these assumptions are necessary consequences of a 
  consistent unification of quantum theory with gravity and spacetime.
  It is fair to say that approaches 
  such as string theory have so far failed to lead to a complete 
  and well defined theory\footnote{A detailed list of results from string theory is contained in
\cite{howfar}.}. Nor have they led to many results that concern 
  the behavior of quantum spacetime at the Planck scale. There is a widely 
  held view that, to give a deeper picture of quantum spacetime, 
  string (or perhaps $\mathcal M$) theory require a background 
  independent formulation.  It is natural to try to construct such a theory using the
methods of loop quantum gravity, some preliminary results in this direction are
given in \cite{11d,Mlee}.
  
  There are several different approaches to a background independent 
  quantum theory of gravity. Besides loop quantum gravity, others include 
  dynamical triangulations\cite{dynamical,AL}, causal sets\cite{causalsets}
and the Gambini-Pullin discrete quantization approach to quantum gravity\cite{GP-discrete}. 
Although these are independently motivated research programs, some of their results
are relevant for loop quantum gravity, because they concern models
which can be understood as arising from spin foam
models by 
simplifications which eliminate certain structures. 

It is also the case that while most results concern quantum general relativity, the
methods of loop quantum gravity  can be applied to construct a large 
  number of different background independent quantum theories, and it can easily incorporate 
  speculative hypotheses about the dimension of spacetime, the form 
  of the action   or the 
  presence of additional symmetries such as supersymmetry.  All of 
  these theories incorporate the basic principles 
of  background independence and diffeomorphism invariance, 
  so that the spacetime geometry is completely dynamical.  
  
  At the same time, by showing that a consistent diffeomorphism 
  invariant quantum field theory can be constructed that represents 
  general relativity, coupled to arbitrary matter fields, in the 
  $3+1$ dimensional world we observe, the success of loop quantum 
  gravity undermines the claim that supersymmetry, strings or higher dimensions 
  are necessary for a consistent quantum theory of gravity.

\section{The basic physical picture}

Before listing the main results of the program, I provide here a simple 
physical picture of how quantum spacetime is understood in loop quantum gravity.
The picture is simple and intuitive, and I will present it without 
the calculations and theorems that back it up. The reader should be 
aware that there is a completely rigorous mathematical formulation supporting this
picture, described in references to the results listed in the next section.  

\subsection{The basics of quantum spacetime}

We begin with the basic picture, which is very simple to state.  To define
the kinematics of a loop quantum gravity theory, pick one from each of the following

\begin{itemize}

\item{}A topological manifold, $\Sigma$, say  $S^3$.  

\item{}A Lie algebra or, more generally, a Hopf algebra or superalgebra, $\cal A$,
say $SU(2)$. 

\end{itemize}

An $\cal A$-spin network, $\Gamma$,  is a graph, whose edges are labeled
by representations of $\cal A$. Each node is labeled by an invariant
in the product of the representations of the incident edges.  

Let $\{ \Gamma \}$ be an embedding of the graph $\Gamma$ into the
manifold $\Sigma$, up to topology.  
Then the Hilbert space ${\cal H}$ of the theory
is defined by the following statement

{\bf $\cal H$ has an orthonormal basis $|\{ \Gamma \} >$ labeled by
the embeddings of the spin networks in the manifold $\Sigma$.}  

The states are interpreted in terms of operators which measure geometric
properties such as  the volumes and areas of regions and surfaces in 
$\Sigma$. Roughly speaking, the spin
network states, $|\{ \Gamma \} >$,  are eigenstates of these observables. 
Each node of a spin network 
contributes a quanta of volume to any region which contains it, which is finite in Planck units and  a function of
the labelings. Each edge similarly contributes a quanta of area to any surface
it crosses. 

There are a number of variants, depending on the specification of 
$\Sigma$. We may specify that $\Sigma$ is a differential manifold, in which
case the classes $\{ \Gamma \} $ are equivalent up to diffeomorphisms
(or, in some formulations, piecewise diffeomorphisms) of $\Sigma$.  
If $\Sigma$ has a boundary, that can indicate an asymptotic region, 
or a horizon of a black hole. 
Or
one can drop $\Sigma$  and build
the theory just from combinatorial graphs. 

The dynamics is given in terms of the basis states as follows.
Consider a small number of local moves, by which one or a few connected nodes change. Give
each an amplitude, which is a function of the labels involved. 

Then, a quantum history from an initial basis state to a final basis state
is a succession of such moves.  Each history has an amplitude which is a product
of the amplitudes of each move.  The quantum theory can then be defined
by a sum over histories. 

Each such history may be considered a quantum version of a spacetime.
It has a discrete causal structure, because the moves can be given 
a partial order which indicate that a later move acted on a region which
resulted from a previous move.

\subsection{Why loops?  The kinematics of diffeomorphism invariant quantum gauge theories}

Where does the picture I just sketched come from?  It has been derived from 
 a new kind 
of quantum gauge theory which is diffeomorphism invariant.  
This construction has been successfully applied to a wide variety of theories, including 
topological field theories, general relativity and 
supergravity. It works in any dimension and all the standard kinds of
matter fields can be included.   The basic 
ideas behind this construction are very simple, from a physics point 
of view, although the math needed to realize them rigorously is a bit 
more involved. 

We begin  by  considering a theory of a 
connection\footnote{All fields described here are forms, the form 
indices are $a=1,...,d$, while the indices $i,j...$ are 
valued in the Lie algebra, $\cal A$},  $A^i_a$, valued in a Lie algebra (or superalgebra)  
$\cal A$ on a $d+1$ dimensional spacetime manifold $\cal M$.   The manifold has
no fixed metric defined on it, fixed or dynamical, all that is fixed is the topology
and differential structure of $\cal M$.  
To quantize a spacetime theory using Hamiltonian methods, we assume that
${\cal M} = \Sigma \times R$, where $\Sigma$ proscribes  the  topology and
differential structure of what we will call ``space".    

We begin with the Hamiltonian theory.  The initial configuration space $\cal C$ will
consist of the possible configurations of the gauge field $A^i_a$ on $\Sigma$. 
To this must be added the conjugate momenta, which are represented by 
the electric field  $\tilde{E}^a_i$, which 
is a vector {\it density}  on $\Sigma$, valued in the lie algebra $G$.  Together, they
coordinatize the phase space, on which is defined
the Poisson bracket relations,
\f
\{ A_a^i  (x) , \tilde{E}^b_j  (y) \} = \delta^3 (x,y) \delta_a^b \delta_j^i
\label{ccr}
\ff

We note that because $\tilde{E}^a_i$ is a density this is well defined 
in the absence of a background metric. 

The physical configuration space consists of 
 equivalence classes of $\cal C$ under the action of  the gauge symmetries
of the theory. These include 
ordinary gauge transformations, with gauge group ${\cal A}(\Sigma )$ 
and the diffeomorphisms of $\Sigma$, denoted, $Diff (\Sigma )$. 

The physical configuration space we are interested in is then,
\f
{\cal C}^{diffeo} = \frac{\mbox{{\cal A}- connections on}  \ \Sigma}{{\cal A}(\Sigma ) 
\times Diff (\Sigma )}
\ff

The problem we want to solve is how to write a corresponding Hilbert 
space of gauge and diffeomorphism invariant states. We proceed in three 
steps. 

\begin{enumerate}

\item{} We construct a Hilbert space of gauge invariant states, called
${\cal H}^{kin}$, on which the diffeomorphisms of $\Sigma$ act {\it 
unitarily} and without anomalies.  

\item{} We mod out by the unitary 
action of the diffeomorphisms, to find the subspace 
\f
{\cal H}^{diffeo} \subset {\cal H}^{kin}
\ff
of diffeomorphism invariant states.  

\item{} We endow ${\cal H}^{diffeo}$ with
an inner product, making it a Hilbert space. 

\end{enumerate}

For the first step, we cannot take the Fock space, which is the usual
starting point for quantization of field theories. The reason is that Fock space depends
on a background metric, and this prevents the construction of a unitary, anomaly
free realization of the diffeomorphism group.  

How do we find an alternative to Fock space, on which we can construct
a unitary, anomaly free action of the diffeomorphisms?   The key point is
that there is no way to do this based on a quantization of the
canonical commutation relations (\ref{ccr}).  Instead we start with
a different set of Poisson bracket relations, made by certain
extended objects.  

An interesting fact about non-Abelian gauge theories is that you cannot
coordinatize the space of connections, mod gauge transformations, completely in terms of $A_a^i$ and
its derivatives at a point\cite{wu-connections}.  To label all distinct
gauge invariant configurations you have to give the values of the 
holonomies. These are defined as follows.   
Given a path  $\alpha \in \Sigma$ the path ordered
exponential is defined as 
\f
U[\alpha ] = P e^{\int_\alpha   A}
\ff
Here $A= A^i\tau_i$, where $\tau_i$ is a generator of 
$\cal A$ in the fundamental representation. 

The  holonomy 
is  the path ordered exponential around a closed loop, $\gamma$
\f
h[\gamma, A] =  P e^{\int_\gamma A}
\ff
This transforms under gauge transformations acting at the beginning of the
loop.  A gauge invariant observable is the Wilson loop, 
which is the trace of an holonomy, 
\f
T[\gamma, A] = Tr P e^{\int_\gamma A}
\ff
To define a conjugate variable, let us for the moment fix\footnote{The construction we are about to sketch works for all $d \geq 2$.}  $d=3$.  The electric
field, as a density is  equivalent to a $2-$form
\f
E_{ab}^i = \epsilon_{abc}\tilde{E}^{c i}
\ff 
This can be integrated against any surface $S \in \Sigma$ to yield
the {\it electric flux through $S$},
\f
E(S)^i = \int_S d^2 S^{ab} E_{ab}^i
\ff
These have a closed algebra under Poisson brackets
\f
\{ U[\alpha , A] , E[S]^i \} = Int [\alpha, S] \  U[\alpha_1 , A]  \tau^i U[\alpha_2 , A]
\label{nccr}
\ff
where $Int [\alpha, S]$ is the intersection number of the surface and 
path, $\alpha_1$ is the path from the origin to the point of intersection,
and $\alpha_2$ is the rest of the path. 

We can now state the key theorems of the subject:

\begin{itemize}

\item{} {\bf Existence:}  There exists a representation of (\ref{nccr}) which
carries a unitary and anomaly-free representation of the group
of diffeomorphisms of $\Sigma$. 

\item{}{\bf Uniqueness: } This representation is unique\cite{unique}\footnote{Modulo some technical assumptions, described below and in \cite{unique}}.  

\end{itemize}

We then take for ${\cal H}^{kin}$ this unique representation.

The mathematics involved to prove these theorems is rather involved.
But the physical picture behind them is very intuitive, and familiar to
physicists. It is in fact the old idea of   duality in Yang-Mills theories: 

{\it The vacuum 
is a dual superconductor, so that the electric flux measured by
$E[S]$ is quantized. The  
physical excitations of this vacuum are defined by 
Wilson loops acting to create normalizable states, which are
states of quantized non-Abelian electric flux.}  

It is easy to give an intuitive picture of how the unique representation
${\cal H}^{kin}$ is built. 

We may begin with a vacuum state $|0>$ which is an eigenstate
of electric flux with vanishing flux everywhere.
\f
E[S]^i  |0> =0
\ff
for all surfaces $S$.  We then create a state which has
one unit of non-abelian electric flux around a loop $\gamma$
by acting on the vacuum with a Wilson loop operator. 
\f
 \hat{T}[\gamma , A] |0> =   |\gamma > 
\label{naive1}
\ff
It pays to work with a gauge invariant version of the electric flux
\f
{\cal E}[S] = \int_S \sqrt{E^i E^i }
\ff
It turns out that this operator can be defined through a regularization
procedure, without breaking diffeomorphism invariance, and the result
is a finite operator. 

The state with one loop excited is an eigenstate of ${\cal E}[S] $
\f
{\cal E}[S] |\gamma > = \hbar Int [\gamma , S] |\gamma > 
\ff
 
If we write the dependence on the connection $A_a^i$ then the
state created by the Wilson loop is just
\f
<A|\gamma > = T[\gamma , A] 
\ff

Given a set of loops $\gamma_i$, $i=1,...M$, for $M$ finite, we can
build up complicated states
\f
 \prod_i \hat{T}[\gamma_i , A] |0> =|  \{   \gamma_i \}  >
\label{naive2}
\ff
Its value is 
\f
<A | \{ \gamma_i \} > = \prod_i T[\gamma_i , A] 
\ff

Such states are not normalizable in Fock space. But this is of little 
concern for us, as there can be no Fock space in the absence of a 
background metric.  It is also the case that there are no operators in the theory
that represents $A^i_a (x)$ or their field strengths $F_{ab}^i (x)$. 
The gauge field is represented only by the non-local operators 
$\hat{T}[\gamma , A] $, but the limits for very small loops, which classically 
would give the field strengths,  do not exist. 

In a certain sense, this representation is closer to the Hilbert spaces of
lattice gauge theory than to Fock space.    Of 
course, a lattice is a background structure, and we don't use a fixed 
lattice here, we simply consider all such states. The new representation
can be thought of roughly as the direct sum of the Hilbert spaces of 
all lattice gauge theories, with 
all possible lattices.  This indeed is how the construction is done in
the rigorous approach.

The key idea is that quantum geometries are built up from such 
states.  There is a translations between gauge fields and 
gravity, which follows from observations III and IV above.   In $3+1$ dimensions the
correspondence gives
${\cal A}=SU(2)$ or $SO(3)$\footnote{For $N=1$ supergravity, $G= Osp(1,2)$.} which 
we will assume for the following.
This gives us  correspondences between flux of the electric field
$\tilde{E}^a_i$ and geometric quantities. Given a surface $S$, it 
turns out that
\f
\mbox{Area of } \ S =  \hbar G \  {\cal E}[S]
\label{correspondence}
\ff
Hence the states created by the Wilson loops acting on the vacuum
$|0>$ 
are eigenstates  of the operators
that measure the areas of surfaces $S$.
So it follows that the dual superconductor picture leads to a 
quantization of areas.  

Note the $\hbar G$ in  (\ref{correspondence}). This 
is necessary because area and electric flux have different dimensions. 
When the correspondence is worked out in detail, they are there 
because   $\hbar$ and $G$ are parameters of a quantum theory 
of gravity. Because of them, flux quanta turn into quanta of area, 
with a minimal quanta of area given by the Planck area $\hbar G$. 

Further, given a region $R \in \Sigma$ the {\it volume of $R$}  can be
expressed as 
\f
\mbox{Volume of} R = (\hbar G)^{\frac{3}{2}}
\int_R  \sqrt{ |det( \tilde{E}^a_i ) |} 
\label{naivevolume}
\ff

It is then of interest to construct simultaneous eigenstates of the operators that 
measure the volumes of all regions $R \in \Sigma$, and the areas of all surfaces
that separate the regions.  This is done by 
combining loops into the spin network graphs introduced above.  
This is necessary because it can be 
shown that after suitable regularization, the operator corresponding to 
(\ref{naivevolume}) annihilates states of the form of (\ref{naive2})
unless there are points where at least two loops intersect. 
Evidently, volume is a property associated with intersections of 
loops. The eigenstates turn out to be spin network states.  

To give the definition of spin network states we return to general $\cal A$. 
A spin network $\Gamma$ is a graph whose edges are labeled with 
irreducible representations of $\cal A$.  The nodes are also labeled by 
invariants (or intertwiners). If a node $n$ has edges incident on it 
with labels $i,j,k,l$, than the node has to be labeled by an
invariant, defined by a map,
\f
\mu : i \otimes j \otimes k \otimes l \rightarrow Id
\ff
For each set of labels on the incident edges there is a finite 
dimensional space of such invariants. We require that there be a 
non-trivial such invariant at each node. 

Given such a spin network $\Gamma $ we can define a {\it spin network 
state},$ |\Gamma > $.  One way to define it is in terms of connections
\f
T [\Gamma, A] = <A |\Gamma > 
\ff
is a generalization of a Wilson 
loop. It is gotten by writing the parallel 
transports of the gauge field $A$ for each edge, in the representation
labeling that edge, and then tracing them together, using the 
invariants labeling each node, to get a gauge invariant functional of 
the connection associated to the whole graph.  

We can also write a spin network state as a sum  of loop
states. This is done by decomposing the representation on each edge
as a product of fundamental representations, and then multiplying out
the formulas for the invariants on nodes in terms of these products, 
and expanding.  

The inner product of ${\cal H}^{kin}$ is chosen so that the spin network states 
comprise an orthonormal basis
\f
<\Gamma |\Gamma^\prime > = \delta_{\Gamma \Gamma^\prime}
\ff

This is of course natural, as such states are eigenstates of the operators 
that measure areas and
volumes.

This completes the construction of the kinematical hilbert space
${\cal H}^{kin}$.  

The next step is to construct  a unitary realization of the diffeomorphism group, 
acting on ${\cal H}^{kin}$.  The rigorous theorems tell us this exists.
At an intuitive level the construction proceeds as follows.   
Given a diffeomorphism $\phi \in Diff (\Sigma )$  we represent it as
a unitary operator as
\f
\hat{U}(\phi ) \circ |\Gamma > = |\phi^{-1} \circ \Gamma >,  \forall \phi \in Diff( \Sigma ) 
\ff
It is trivial to check the unitarity. This follows from the fact that for any spin network
$\Gamma$ and any diffeomorphism $\phi \in Diff( \Sigma )$,  $|\Gamma >$ and
$| \phi \circ \Gamma >$ are both normalizable states, with the same norm.   

It is less trivial, but still true, to prove that there is no anomaly.  

The next step is to change basis from the $<A| $ basis
in which states are represented as functionals $\Psi (A)= <A|\Psi >$ 
 to the spin network
basis where they are functionals $\Psi (\Gamma ) = <\Gamma |\Psi >$.
Given that the $<\Gamma |$ comprise an orthogonal basis, this
can be done rigorously.  The functional transform 
\f
\tilde{\Psi} (\Gamma ) = \int d\mu (A) T[\Gamma , A] \Psi (A)
\ff
is known precisely, the measure required is called the
Ashtekar-Lewandowski measure.  Heuristically this is analogous to the Fourier
transform as $<A|\Gamma > = T[\Gamma , A]$ gives a basis of states that
the Hamiltonian acts simply on.  Indeed, the key discovery that makes the
quantization of general relativity 
 possible is that the Hamiltonian and diffeomorphism constraints,
which make up the hamiltonian, take any state $ T[\Gamma , A]$ to a state
of the same form, but with a different $\Gamma$.  This is analogous to the fact
that the Fourier transform is useful in ordinary quantum mechanics because the
hamiltonian acts algebraically on momentum eigenstates. 

One then defines a space of diffeomorphism invariant states
\f
{\cal H}^{diffeo} \subset {\cal H}^{kin}
\ff 
containing all states for which $\Psi [\phi \circ \Gamma ] = \Psi [\Gamma ]$,
for all $\phi \in Diff( \Sigma )$.  It is easy to  construct such states. For example,
let ${\cal K}$ be any knot or graph invariant, then 
\f
\Psi_{\cal K}[\Gamma ]={\cal K}[\Gamma ]
\ff
is in ${\cal H}^{diffeo}$.    Similarly, let $\{ \Gamma \} $ be the diffeomorphism 
equivalence class of
the network $\Gamma $. Then the characteristic state defined by 
\f
\Psi_{\{ \Gamma \} } [\Gamma^\prime  ] = 1 \  \mbox{if}  \  \Gamma^\prime \in \{ \Gamma \} 
\ff
and zero otherwise is in ${\cal H}^{diffeo}$. In fact, it can be shown that these
provide an orthonormal basis of ${\cal H}^{diffeo}$.  

Thus, the diffeomorphism invariant states perfectly combine the principles of
duality and diffeomorphism invariance.  Each state gives an amplitude to
diffeomorphism classes of collections of Wilson loops.  These states have
physical meaning, given to them by the fact that they diagonalize diffeomorphism
invariant observables that measure the geometry of the 
spatial slice. Examples of these which we mentioned 
above are the volume of the universe and the area of its boundary.  

There is an unusual feature of the construction, worth commenting on.
In a certain sense the kinematical Hilbert space, ${\cal H}^{kin}$ has
too many states to represent a sensible physical theory. This is because
any two spin network basis states are orthogonal, no matter by how little
the graphs on which they are constructed differ.  As a consequence it
can be shown that there is no countable basis, so that the space is
{\it non-seperable}.  

This would be a disaster for physics. But we do not do physics in this space,
we do physics in ${\cal H}^{diffeo}$, which is much smaller because we
have identified all graphs that differ by a diffeomorphism.  It can be shown
that this space is seperable\footnote{if one mods out by piecewise smooth
diffeomorphisms.}.  

To complete the theory one has to represent   the dynamical evolution equations  
as operators on  these 
diffeomorphism invariant states.   
This can be done in particular examples such as general relativity coupled to arbitrary 
matter fields in $3+1$ dimensions and supergravity.   Exact expressions for the 
dynamical evolution of these states is known in both hamiltonian and path integral form. 
The action changes the graphs by local rules.  A typical action is to act at a trivalent node, converting it to a triangle. 


From the quantum Einstein equations we have exact 
closed form expressions for the 
amplitudes for these processes. This then 
gives rise to the picture of
dynamics we sketched above in section 3.1.

We now turn to the methods by which the evolution amplitudes are constructed. 

\subsection{Dynamics of constrained topological field theories}

Observation {\bf IV} means in details that all gravitational theories of interest, 
including general relativity and supergravity in 4 dimensions, can be described  by an action, 
which is generically of the form\footnote{This kind of formulation of general
relativity was first discovered by Plebanksi\cite{plebanski} and
later independently by \cite{JSS,CDJ}. The corresponding simplification
of the Hamiltonian theory was independently 
discovered by Sen\cite{sen} and formalized by Ashtekar\cite{abhay}.  By now several
 different connections are used in loop quantum gravity. These include the
 self-dual part of the spacetime connection\cite{sen,abhay}, and a real
 SU(2) connection introduced by Barbero\cite{barbero} and exploited by
 Thiemann\cite{thomas}. There are also alternate formulations that
 use both the left and right handed parts of the spacetime connection,
 \cite{yime-holo,me-holo}.},
\f S=S^{topological} + S^{constraints} + S^{matter}
\label{master}
\ff
To describe the detailed form, its simplest first to fix the dimension to be four,
in which case $G=SU(2)$. 

The first term describes a topological theory called $BF$ theory.  It 
depends on a $2$ form $B^i$ and the field strength $F^i$ of  a connection, $A^i$,
all valued in a Lie algebra of  $SU(2)$.  Thus, $i=1,2,3$.   The action is, 
\f
S^{topological}=\int_{\cal M}   ( B^i \wedge F_i - \frac{\Lambda}{2}B^i \wedge B_i )
\label{topological}
\ff
The field equations which follow are
\f
F^i = \Lambda B^i
\label{Ftop}
\ff
\f
{\cal D} \wedge B^i =0 
\label{Btop}
\ff
where $\cal D$ is the $SU(2)$ gauge covariant derivative.

The second term contains  a quadratic function of $B^i$, which 
can be  expressed as
\f
S^{constraint} = \int_{\cal M} \phi_{ij} B^i \wedge B^j
\ff
where $\phi_{ij}$ is a symmetric, traceless matrix of scalar fields. 
Variation of the independent components of $\phi$ produces a quadratic equation in $B^i$ whose solution 
turns the theory into
general relativity.   These are
\f
B^i \wedge B^j = \frac{1}{3}\delta^{ij}B^k \wedge B_k
\label{quad}
\ff
The solutions to this are all of the following form:  there exists a frame field $e_a^I$ for 
$I=0,...,3=0,i$, such that
the $B^i$ are the self-dual two forms of the metric associated to $e_a^I$.  That is,
\f
B^i = e^0 \wedge e^i + (\imath) \epsilon^{ijk} e_j \wedge e_k
\ff
where the $\imath$ is there for the Lorentzian case and not for the Euclidean case. 
Thus, the metric is not fundamental, instead the frame field appears when we solve
the equations, (\ref{quad}),  that constrain the degrees of freedom of the topological field theory. 

The third term contains coupling to matter fields, such as spinors, scalars and Yang-Mills fields. 
Interestingly enough, these can be written in terms of the fields involved
in the other two terms\cite{CDJ}. 

It turns out that the same trick works in all dimensions, where $B^i$ is now a $d-2$ form. 
For $d=2+1$, the $BF$ theory is equivalent to general relativity\cite{2+1}. For spatial
dimension $d>3$ the extension has been given in \cite{higher}.  It is also known
how to express supergravity in $d=3+1$ \cite{super} and $d=10+1$ \cite{11d}
as constrained topological field theories.  

Now we can return to the canonical quantization, and discover the structure
we assumed above.  Given a choice of time 
coordinate, $t$, which represents $\Sigma$ as constant time 
slicings\footnote{However, among the gauge symmetries of the theory are diffeomorphisms of 
$\cal M$ that take any ``spatial" slice representing  $\Sigma$ to any other slice. 
If the theory implements the quantum version of the constraint that generates this 
gauge symmetry, it will not depend on the choice of slice used.}, we can find the 
canonical momenta to $A^i_a$.  We see from the form of the action that the only time 
derivatives are in the first term, 
\f
S^{topological}=\int dt \int_{\Sigma}   ( B^i \wedge \dot{A}_i + ...
\ff
Hence the canonical momenta to the gauge field are contained 
in conjugate electric fields, which can be expressed in terms of the pull back
of the two forms of our theory to the spatial manifold  $\Sigma$. 
\f
E^a_i = \epsilon^{abc} B_{bc i}
\label{EB}
\ff
where the  $\epsilon^{abc}$ is defined on the spatial manifold $\Sigma$, 
making $E^i$ a vector density on $\Sigma$.  
When the quadratic constraints from $S^{constraints}$ are solved, all the 
metric information is contained in the $B^i$'s, and hence 
is represented by the electric field $E^a_i$ conjugate to the gauge field $A^i$.   
The gauge field $A^i_a$ will turn out to code components of the spacetime connection.
This is a reversal of the older $ADM$ way of understanding the dynamics of the 
gravitational field, but it turns out to be deeper, and much more progress can be made with it.  

We can easily see how the restriction from the topological $BF$ theory to
general relativity by a quadratic equation works in the hamiltonian formulation.   
As the field equations of the $BF$ theory, (\ref{Ftop},\ref{Btop}) are expressed as 
spacetime forms, 
they pull back to equations in the three surface $\Sigma$. These  must
hold the canonical theory.  The covariant conservation of $B^i$
given by (\ref{Btop}) pulls back, given (\ref{EB}), to Gauss's law
\f
{\cal G}^i={\cal  D}_a \tilde{E}^{ai}=0.
\label{gauss}
\ff
As in any Yang-Mills gauge theory, these are first class constraints that generate the ordinary
gauge transformations.  The field equation of the topological field theory
(\ref{Ftop}) pulls back to
\f
{\cal F}_{ab}^i = F_{ab}^i + \Lambda \epsilon_{abc} E^{c i} =0
\label{topconstraints}
\ff
These are constraints which tell us that the curvature is entirely determined by the $\tilde{E}^a_i$.
Hence,  the connection has no local independent degrees of freedom.

To give local dynamics to the connection we want to impose conditions on the fields, 
which will restrict the number of independent constraints.  As mentioned above, 
to get general relativity in any dimension, it suffices to impose quadratic  
conditions on the $B^i$'s.  In fact, it is easy to see how this is realized in the
Hamiltonian form of the theory, for the case of four spacetime dimensions. 

The required quadratic constraints are equations of motion, hence they should
be generated by the hamiltonian. This suggests that a hamiltonian at most cubic
in fields should suffice.  
However in spacetime diffeomorphism invariant theories the hamiltonian must be a linear
combination of constraints, so as to avoid any preferred time coordinates.  
On general grounds we expect there must be four more constraints, to generate
the four dimensional diffeomorphisms.  

These constraints
should come from restricting the constraints (\ref{topconstraints}) that give 
the topological field theory. The simplest way to do this turns out
to lead to general relativity\footnote{The same reasoning in the supersymmetric
case leads to supergravity}.  First, we can trace (\ref{topconstraints})
with one power of the momenta, to find,
\f
D_a = {\cal F}_{ab}^i  E_i^b = F_{ab}^i  E^b_i =0
\label{diffconstraint}
\ff
It is easy to show that these generate
the diffeomorphisms of $\Sigma$.   The next simplest thing to do is to
trace (\ref{topconstraints}) with two powers of the momenta. This yields
the desired cubic constraint, 
\f
{\cal H} = \epsilon_{ijk }E^{aj} E^{bk}  {\cal F}_{ab}^i  = \epsilon_{ijk } E^{aj} E^{bk}  [F_{ab}^i + 
\Lambda \epsilon_{abc} E^{c i}] =0
\label{hamiltonianconstraint}
\ff
This is the Hamiltonian constraint. Remarkably, unlike older approaches, it is polynomial in the fields. This makes it possible to translate it into a quantum operator using non-perturbative
methods\footnote{The form (\ref{hamiltonianconstraint})  holds for the Ashtekar-Sen variables, 
in which, for the Lorentzian case, $A_a^i$ is a complex variable. One can also work
with the real Barbero variable, at a cost of an additional term added to the Hamiltonian
constraint in the Lorentzian theory. However, it is still possible to convert
the Hamiltonian constraint to a finite, well defined operator in this case.}.  

Can one stop here? One can if the constraints form a closed algebra. It turns out
that the system of 7 constraints, consisting of (\ref{gauss}),(\ref{diffconstraint})
and (\ref{hamiltonianconstraint}) do form a closed first class algebra.  
Hence it is consistent to impose only these four of the
nine equations in (\ref{topconstraints}). The result must be a theory with local
degrees of freedom\footnote{To count degrees of freedom, note that in the topological
field theory Gauss's law (\ref{gauss}) follows from (\ref{topconstraints}), but this
is not true of the restricted set (\ref{diffconstraint},\ref{hamiltonianconstraint}). Hence
there is a net of two ($9-7$ )fewer equations, so there are two physical degrees 
of freedom. These are the polarizations of gravitational waves.}
It is not hard to show that it is general relativity.  It is also straightforward
to show that for $d=4$   the 
constraints (\ref{gauss},\ref{diffconstraint},\ref{hamiltonianconstraint}) do follows
from the action (\ref{master}).  

\subsection{Horizons, black holes and boundaries}

Loop quantum gravity has led to a number of important results about black holes
and cosmological horizons.  In most of these results  a horizon is modeled
as a boundary of spacetime.  Conditions are imposed at the 
boundary which capture the physical
feature that at a horizon nulls rays neither diverge nor converge. 
As a result, only the degrees of freedom on the horizon and its exterior are
quantized.  This method certainly limits the kinds of questions that can be asked
about black holes but, within this limitation, there is a complete description of the
states associated with the black hole horizon.  

These results are made possible by the relationship with topological field theory. 
The reason is easy to describe. Let us now assume that the spatial manifold $\Sigma$ has
a boundary $\partial \Sigma = {\cal B}$ which is a closed, oriented two dimensional
surface.  For the classical and quantum dynamics to be well defined we must
fix boundary conditions at $\cal B$ and we must also add a boundary term to the
action of the theory.  There is a very natural class of boundary conditions which
follow from the relation to topological field theory. The reason is that the only
derivatives of the action of general relativity are shared with the  topological
field theory. We can make use of natural relationships between topological field
theories of different dimension, called the ladder of dimensions\footnote{Whose relevance
for quantum gravity was first emphasized by Louis Crane.}.  
This works in any dimension, but there is an especially nice 
situation in $3+1$ dimension because the natural boundary theory associated
with a $BF$ theory is a Chern-Simons theory on the three dimensional boundary
of spacetime.  To realize this we add a boundary term to (\ref{master}), given by
\f
S^{boundary} = {k \over 4\pi} \int_{{\cal B}\times R} Y_{CS} (A)
\ff
where $Y_{CS} (A)$ is the Chern-Simons three form of the connection pulled
back to the spacetime boundary.  One can then show that the classical and
quantum dynamics is well satisfied, so long as a boundary condition is
satisfied on $\cal B$, which is\cite{linking}
\f
B^i|_{\cal B} = {k \over 2\pi }F^i|_{\cal B} 
\label{bc}
\ff

It turns out that this condition precisely characterizes black hole horizons, with
the constant $k$ related to the surface gravity
at the horizon\cite{isolated}. it applies also to cosmological horizons\cite{linking} and to  timelike  
boundary conditions in the case of
a non-zero cosmological constant, of either sign\cite{me-holo}. In this case the consistency
of the boundary condition (\ref{bc}) with the constraint (\ref{hamiltonianconstraint}) 
implies that,
\f
k = {6\pi \over G\hbar \Lambda} .  
\ff
Since Chern-Simons theory is precisely understood, this leads to a detailed
understanding of the physics at these boundaries, including horizons\cite{linking}-\cite{me-holo}.

\subsection{The path integral formulation: spin foam}

Path integral formulations of loop quantum gravity are known as spin foam models.
There are several different  constructions of them, and up to 
technical issues\footnote{For the interested reader these are described below.},
they yield the same class of theories. Each construction yields a spacetime
history, or spin foam, $\cal F$ which evolves an initial spin network state
$|\Gamma_{in}> $ into   a final state $| \Gamma_{out}> $.  A spin foam
history $\cal F$,  is
itself a combinatorial structure, whose boundary is the initial and final
state
\f
\partial {\cal F} = \Gamma_{out}-\Gamma_{in}
\ff
Here are several equivalent characterizations of a history, $\cal F$.

\begin{itemize}

\item{}{\bf A spin foam as a causal history.}  A {\it causal} spin foam is a succession of
local moves $m_i$ that transform the initial state $\Gamma_{int}$ to
the final state $\Gamma_{out}$\cite{F-foam}-\cite{fmls1}.  The moves $m_i$ are considered the
discrete analogues of the {\it events} of a continuous spacetime. They 
have a partial order,
which give the history a discrete analogue of the causal structure of 
a Lorentzian spacetime.  Consequently, each causal spin foam history
is endowed with discrete analogues of structures such as light cones and horizons.

\item{}{\bf A spin foam as a one dimension higher analogue of a Feynman diagram}
in which spin networks play the role normally played by incoming and 
outgoing particle states\cite{RR-foam}. 
Propagators for edges are two dimensional 
spacetime surfaces, they meet along edges which are propagators for nodes.
Each surface is labeled by a representation, as the edge it propagates. Similarly,
each edge of the spin foam is labeled by an invariant corresponding to the node
of the spin network it propagates. 
Interactions occur at vertices in the foam, where several edges meet.
 Any cross section of such a spin foam is
a spin network.  The interactions are events where the spin network changes by
a local move.  

\item{}{\bf Spin foams as Feynman diagrams of a matrix model.}  The analogy
to Feynman diagrams is made precise by showing that there for every spin foam model
there is a matrix model (more precisely a dynamical system on a group manifold) 
such that the spin foams are the Feynman diagrams of
that matrix model\cite{foam-matrix}.  Each foam is then a Feynman diagram, and like in ordinary
quantum field theory, there are sums over intermediate states on internal
propagators. In the spin foam case these are sums over representation labels
on two surfaces and sums over invariant labels on edges.  But unlike ordinary $QFT$,
in a large class of physically relevant models, the sums over intermediate states are
finite.  

\item{}{\bf A spin foam as a triangulation of spacetime}  For simplicity, consider the
case where the initial and final spin network are four valent graphs. They then are
each dual to a three dimensional 
triangulation \footnote{Technically, a pseudo-manifold,} in which each node is dual to 
tetrahedra, whose four faces are dual to the four edges incident on that node\cite{mike-foam,BC}.
Each face of the dual triangulation is now labeled by a representation, while each
tetrahedron is labeled by an invariant.  
There are then spin foams constructed from a four dimensional triangulation,
whose boundary is the three triangulation dual to the initial and final spin network.
Each four simplex is dual to an event in the previous characterizations of a 
spin foam.  

\end{itemize}

Following the rules of quantum theory, the dynamics of the theory is specified
when we have assigned an amplitude to each history.  
The amplitude of each foam ${\cal A}[{\cal F}]$ is given by a product
\f
{\cal A}[{\cal F}]= {\cal N}^{-1} \prod_{events} {\cal A}[event ]
\ff
of factors for each event (or vertex or four-simplex), where each amplitude is
a function of the labels of surfaces and edges incident at that event,.  
$\cal N$ is a normalization factor which   depends on the labels
on lower dimensional structures.   Several models have been 
studied as candidates for quantum general relativity. The best studied
is the Barrett-Crane model, which is derived by following a strategy which
is naturally indicated by the realization that general relativity is a constrained
topological field theory. The first step is to define a spin foam model which exactly realizes
the quantization of the $BF$ theory given in (\ref{topological}).  

The second step is to implement 
the quadratic constraint, (\ref{quad}),  on the sums over labels in the path integral representation
of the topological field theory. It turns out there is a very beautiful and natural
way to do this. The result is that the path integral measure for quantum general
relativity is {\it exactly the same as the measure of the corresponding topological field theory.}
The only difference is that in quantum general relativity the sum over labels is restricted to 
a subset of representations and invariants. This restriction implements the quadratic
constraint and, by destroying the topological invariance, leads to a theory with local
degrees of freedom. 
This simple construction is known to work
for quantum general relativity in all dimensions, 
for both Lorentzian and Euclidean theories\cite{higher}.

Following the conventional rules of quantum theory, amplitudes of physical interest
are to be constructed from summing the corresponding amplitudes over all 
spin foam histories.  Thus, we have
\f
<\Gamma_{out}|\Gamma_{in}> = \sum_{{\cal F} s.t. \partial {\cal F} =\Gamma_{out}-\Gamma_{in} }
{\cal A}[{\cal F} ]
\ff

A spin foam model then provides a precise realization of the sum over spacetime
histories, conjectured and described formally in early work by Hawking, Hartle and
others.  Given such a model, one can use it to construct different amplitudes of
interest including projection operators onto the space of physical states, annihilated
by all the constraints. One can construct as well physical evolution amplitudes,
such as the amplitude to evolve from the initial state to the final state through
histories with a fixed spacetime volume.

\section{The main results of loop quantum gravity }

We can now turn to listing some of the main results which have been obtained
concerning loop quantum gravity and spin foam models.

\subsection{The fundamental results of the canonical theory}

The first set of results concern the construction of a Hilbert space of states of a gauge field valued
in a Lie algebra or superalgebra  $\cal A$, 
invariant under local gauge transformations and diffeomorphisms of $\Sigma$. 

The setting for the results that follow is  a bare differentiable manifold 
$\Sigma$, with no metric structure, on which is defined the phase 
space of a gauge theory, with gauge group $G$. The only non-dynamical structure that is
fixed is a three manifold $\Sigma$, with a given topology and
differential structure. There are no fixed classical fields such as
metrics, connections or matter fields on $\Sigma$. The only exception 
is in modeling the quantization of spacetime regions with boundary,
as in the asymptotically flat or $AdS$ context, or in the presence of
a black hole or cosmological horizon. In these cases 
fields may be fixed on the boundary $\partial \Sigma$
to represent physical conditions held fixed there. 
The dynamics is formulated in terms of a 
diffeomorphism invariant action which is a functional only of that 
gauge field and its derivatives.  

 {\bf 1.}  The states of the theory are known
precisely\cite{tedlee,loop1,loop2,gangof5,thomas-thesis,Carlo-book}.  
The kinematical Hilbert space ${\cal H}^{kin}$ has been rigorously
constructed and has all the properties described in the last section.    
The Hilbert space ${\cal H}^{diffeo}$ 
of spatially diffeomorphism invariant and gauge invariant states of a gauge field on a manifold $\Sigma$, 
has an orthonormal basis, $|\{ \Gamma \} > $ whose elements  are in one to one correspondence
with the diffeomorphism equivalence classes of embeddings of
 spin networks\cite{sn-roger},
into $\Sigma$ \cite{sn1}.  The inner product is given by 
\f
< \{ \Gamma \} |\{ \Gamma^\prime \} >=  \delta_{\{ \Gamma \} \{ \Gamma^\prime \} }.
\ff
Here $\{ \Gamma \} $ refers to the equivalence class of graphs under piecewise
smooth diffeomorphisms of $\Sigma$.  
In the case of pure
general relativity in $3+1$ dimensions, with vanishing cosmological constant, 
the gauge group is $SU(2)$.  In this case the labels
on the edges are given by ordinary $SU(2)$ spins. For details
see \cite{sn-roger,sn1}.
  
  \blankline

  {\bf 2.}  Certain spatially diffeomorphism invariant 
  observables have been constructed. After a
  suitable regularization procedure these turn out to be
  represented by {\it finite} operators on ${\mathcal H}^{diffeo}$, the space
  of spin network states\cite{spain,volume,volume2}.
  In the case of general relativity and supergravity, these include the volume of the universe,
  the area of the boundary of the universe, or of any surface defined
  by the values of matter fields.     These operators all preserve the diffeomorphism
  invariance of the states\cite{diffeo}.  

Other operators also have
  been constructed, for example an operator that measures
  angles in the quantum geometry\cite{seth-angles} and the lengths of curves in
$\Sigma$\cite{tt-length}.

   \blankline
  
  {\bf 3.}  The area,  volume and length operators have discrete, finite
  spectra, valued in terms of the Planck length\cite{volume,volume2,renate-volume,tt-length}.
  There is hence a
  smallest possible volume,   a smallest possible area, and a smallest
possible length, each of Planck scale. 
 The spectra have been 
computed in \cite{renate-volume,johannes-thomas-volume}.

 \blankline

{\bf 4.}  The area and volume operators can be promoted to genuine
physical observables, by gauge fixing the time gauge so that at least
locally time is measured by a physical field\cite{me-maryland,diffeo}.  
The discrete
spectra remain for such physical observables, hence the spectra of
area and volume constitute genuine physical predictions of the quantum
theory of gravity.

 \blankline

  {\bf 5.}  Among the operators that have been constructed and found to
  be finite on ${\mathcal H}^{diffeo}$ 
  is the Hamiltonian constraint (or, as it is often called,
  the {\it Wheeler de Witt} equation\cite{ham}-\cite{thomas-ham}.)
  Not only can the Wheeler
  deWitt equation be precisely defined, it can be solved exactly.
  Several infinite sets of solutions
  have been constructed, as certain superpositions of the spin
  network basis states, for all values of the cosmological
  constant\cite{loop1,thomas}.
  These are exact, physical states of quantum general relativity.

 {\bf 6} If one fixes a physical
  time coordinate, in terms of the values of some physical fields,
  one can also define the Hamiltonian for evolution in that physical
  time coordinate\cite{me-maryland} and it is also given by a finite
  operator on a suitable extension of ${\mathcal H}^{diffeo}$ including
  matter fields.

   \blankline
  
  {\bf 8.}  There is a rigorous formulation of the Hamiltonian 
  quantization of general relativity in which all the preceding 
  results are reproduced as theorems\cite{gangof5,thomas,thomas-thesis}. 
In the context of this rigorous approach, there is a fundamental uniqueness
theorem\cite{unique}, which essentially says the following\footnote{A more
precise statement of the theorem follows\cite{thomas-personal}.

Let $A(e)$ be the holonomy of a classical connection along a piecewise
analytic path $e$, let $E_f(S)=\int_S Tr(f *E)$ be the electric flux of
the classical two-form $*E$ through the piecewise analytic surface $S$ where
$f $ s a Lie algebra valued smooth scalar field. Let $D$ be the group of
piecewise analytic spatial diffeomorphisms of the hypersurface and
$G$ the group of local gauge transformations. Let
$W$ be the Weyl algebra generated by
$[A(e),A(e')]=0$ and
$W_f(S) A(e) W_f(S)^{-1}=\exp(i\hbar {\cal L}_{\chi_{E_f(S)}})\cdot A(e)$. 

where we have, 
$A(e)^*=A(e^{-1})^T  $ (T=transposition of matrices)
$W_f(S)^*=W_{-f}(S)$

Finally consider group of automorphisms of $W$ labeled by the semidirect
product $G \times D$ given by

$\alpha_g(A(e))=g(b(e)) A(e) g(f(e))^{-1} (g\in G; b(e),f(e)$  beginning and
                                          end point of $e$)
$\alpha_d(A(e))=A(d(e))$

and similar for $W_f(S).$

We are looking for representations $\pi$ of the *algebra $W$ on a Hilbert
space$ H$ which

1. contains a cyclic vector $ \Omega$, i.e the states$ \pi(a)\Omega$ span $H$
as we let $a$ vary through $W.$

2. implements $G |x D$ unitarily by
$U(g,d)\pi(a)\Omega=\pi(\alpha_{g,d}(a))\Omega$  for all $a \in W$

3. is weakly cont. wrt $t \to \pi(W_{tf}(S))$ for all $f, S$ for real parameter
$t$, and that  $\Omega$ is in the common dense domain of all the $\pi(E_f(S)$. 
That means that matrix elements of $\pi(W_{tf}(S)$ become matrix
elements of the unit operator as$ t \to 0$ which by Stone's theorem means
that the flux operators$ \pi(E_f(S))$ exist. This technical assumption
is necessary also in Stone von Neumann's theorem without which the
uniqueness of the Schroedinger rep of QM is wrong.

A different way of saying this is that $(\pi,\Omega,H)$ are the GNS data
for a state (pos. lin. functional) $\omega$ on $W $satisfying the invariance
condition $\omega\circ \alpha_{g,d} =\omega$ and the continuity assumption.

Then the THEOREM is:
{\it There is only one solution, the $AILRS$ Hilbert space.
Moreover, the representation $\pi_{AILRS}$ is irreducible (that is, not only
one vector is cyclic but every vector is cyclic) which excludes the
existence of spurious superselection sectors.} }: 

{\it Consider an
approach to a quantum gauge theory on a $d \geq 2$ dimensional 
manifold without metric. Assume that the Wilson loop 
operator and area (or generally electric flux) operator are well defined on a kinematical
 hilbert space,  ${\cal H}^{kin}$. Assume also that ${\cal H}^{kin}$ 
carries a unitary anomaly free representation
of $Diff(\Sigma )$ so that the space ${\cal H}^{diff}$ of diffeomorphism 
invariant states may be constructed (formally as a subspace of a dual space of  ${\cal H}^{kin}$. 
Then the Hilbert space of the theory is isomorphic to that just described. }
  
  \subsection{Results on path integrals and spin foams}
  
  {\bf 9.}  The dynamics of the spin network states can be expressed
  in a path integral formalism, called
  spin foams\cite{baezfoam}-\cite{alejandro-review},\cite{F-foam}-\cite{BC}\footnote{For the most 
  recent review see \cite{alejandro-review}.}.
  The histories by which spin network
  states evolve to other spin network states, called spin foam
  histories,  are explicitly known.
  A spin foam history is a labeled combinatorial
  structures, which can be described as a branched labeled two complex.
Spin foam models have
  been derived in  several different ways, and the results
  agree as to the general form of a spin foam amplitude. These
include: 1) by exponentiation of the Hamiltonian
  constraint, 2) directly from a discrete approximation
  to the classical spacetime theory, 3)  by constraining the
  summations in a finite state sum formulation of a four dimensional
  topological invariant, 4) from a matrix model on the space of
  fields over the group, 5) by postulating spacetime events are local
  moves in spin networks.

   \blankline
   
  {\bf 10.} Evolution amplitudes corresponding to the
   quantization of the Einstein equations in $3+1$ dimensions,
   are known precisely\cite{BC,alejandro-review}
   for vanishing and non-vanishing values of the cosmological
   constant, and for both the Euclidean and Lorentzian theories.

    \blankline

  {\bf 11.}  The sum over spin foams has two parts, a sum over graphs
  representing histories of spin networks, and, on each, a sum
  over the labels.  The sums over labels are known from both analytic
  and numerical results to be
  convergent\cite{finite-foam,finite-dan} for some spin foam
  models, including some corresponding to the quantization of
  the Einstein equations in $2+1$ and $3+1$ dimensions.

   \blankline
  
 {\bf 12.} For some spin foam model in $2+1$ dimensions, it has been shown that
 the sum over spin foam histories is Borel
  summable\cite{laurent-borel}.

   \blankline

  {\bf 13.} The physical inner product, which is the inner product on solutions
  to all the constraints, has an exact expression, given in terms
  of a summation over spin foam amplitudes\cite{RR-foam}.

   \blankline
  
   {\bf 14.} The spin foam models have been extended to include
  gauge and spinor degrees of freedom.  
  
   \blankline

{\bf 15.} Spin foam models with matter
have been extensively studied in $2+1$ dimensions\cite{laurent-matter}.
It has been shown that the limit $G_N \rightarrow 0$ reproduces
the Feynman diagram expansion of the matter quantum field theory
on $2+1$ dimensional Minkowski spacetime.  For $G >  0$ the
matter quantum field theory on $2+1$ dimensional
$\kappa$-Minkowski spacetime is reproduced.  This shows that at least in 
$2+1$ dimensions the scattering of matter coupled to quantum gravity
is described by a version of deformed special relativity.  This is discussed
more in sections 4.5 and 5.  
  
  {\bf 16. } Spin
  foam models appropriate for Lorentzian quantum gravity, called
  causal spin foams, have quantum analogues of all the basic features
  of general relativistic spacetimes\footnote{For more details on these
  models and the resulting physical picture, see\cite{Fotini-Wheeler}.}. 
  These include
  dynamically generated
  causal structure, light cones and a discrete analogue of
  multifingered time, which is the freedom to slice the spacetime
  many different ways into
  sequences of spatial slices\cite{F-foam}.
  The spatial slices are spin networks,
  which are quantum analogues of spatial geometries\footnote{It should be mentioned
also that there are very impressive results relevant to the spin foam program from
a research program called {\it causal dynamical triangulations.}  This work shows
that Lorentzian and Euclidean path integrals for quantum gravity fall into different
universality classes, with the Lorentzian path integrals being more
 convergent\cite{AL}-\cite{AL3+1}. For the first time there are results that suggest
that a dynamical $3+1$ spacetime may emerge from critical behavior of a
path integral over combinatorially defined spacetimes\cite{AL3+1}.  }.
  
\subsection{Matter and unification}

   \blankline

  {\bf 17.}  Coupling to all the standard forms of matter fields are
  understood,  including gauge 
  fields, spinors, scalars and higher $p$-form gauge fields\footnote{For 
  supergravity, see result 31, below.}\cite{Carlo-book,thomas-thesis}.  
It is known how to extend
  the definition of the spatially diffeomorphism invariant states to
  include all the standard kinds of matter fields, and the 
  corresponding terms for the Hamiltonian constraint are known in 
  closed form, and are finite on the space of diffeomorphism 
  invariant states. 
  These states are also invariant under ordinary Yang-Mills and
$p$-form gauge transformations\footnote{To my knowledge whether 
  loop quantum gravity suffers from the fermion doubling problem is
  an open question.}.  Inclusion of matter fields does not
  affect the finiteness and discreteness of the area and volume
  observables. 


\blankline

  \subsection{Results on black holes and horizons}

  {\bf 18.}  Several kinds of boundaries may be incorporated in the
  theory including timelike boundaries, in the presence of both
  positive and negative cosmological constant, and null boundaries
  such as black hole and cosmological
  horizons\cite{linking}-\cite{isolated,yime-holo,me-holo}.
  In all these cases
  the boundary states and observables are understood in terms of
  structures derived from Chern-Simons theory.  The boundary
 theory  provides a detailed microscopic description of the
 physics of horizons and other boundaries.
  
   \blankline

  {\bf 19.}   The horizon entropy (\ref{bb})
 is completely explained in terms of the
  statistical mechanics of the state spaces associated with the
  degrees of freedom on the horizon. This has been found
  to work for a large class of black holes, including Schwarzschild
  black holes\cite{kirill1,isolated}.  LQG also gives the correct results for
dilatonic black holes\cite{dilatonic}. 
  
  The boundary Hilbert spaces decompose into eigenspaces, one
  for each eigenvalue of the operator that measures the area of the
  boundary\cite{linking}.
  For each area eigenvalue, the Hilbert space is finite
  dimensional. The entropy may be computed and it agrees precisely
  with the Bekenstein-Hawking semiclassical result,
\f
S = {A[S] \over 4 \hbar G_{Newton}}
\label{bb}
\ff

  The calculated entropy is proportional to  a parameter,
  which is called the Imirzi parameter. This can be understood
  either as a free parameter that labels a one parameter family
  of spin network representations, or as the (finite) ratio of the
  bare to renormalized Newton's constant. 

The Imirzi parameter can be calculated by computing the entropy
statistically. To do this one has to make an assumption about which
microstates of the horizon should be counted as corresponding to the
macrostate which is the Schwarzchild black hole.  Two different
answers, both of order one have been gotten depending on which assumption
is made\cite{match,nomatch}

The Imirzi parameter can also be fixed  by an argument invented by Dreyer, involving  quasi normal modes of black holes\cite{olaf-bh}.  The value gotten by this method matches one of the values gotten by calculating the ratio of
microstates  to macrostates explicitly. 

Dreyer's argument depends  on a remarkable and precise
  coincidence between an asymptotic value
  of the quasi normal mode frequency and a number which appears in
  the loop quantum gravity description of horizons. The value of
  the asymptotic quasi normal mode frequency was at first known
  only numerically, but it has been very recently derived
  analytically by Motl\cite{motl}.  Once Dreyer's argument fixes the
  Imirzi parameter, the Bekenstein-Hawking relation (\ref{bb}) is
  predicted exactly for black hole and cosmological
  horizons.
  
   \blankline

  {\bf 20.}  Corrections to the Bekenstein entropy have been calculated and
found to be logarithmic\cite{logcorrect}.

   \blankline

{\bf 21}To study the problem of the fate of the singularity models of the interior of a 
spherically symmetric black
hole horizon have been studied\cite{ modesto,oliverviqarbh,abhaymartinbh}.  These are  $1+1$ dimensional field theories.  It is found that the  black hole singularities are eliminated\cite{ modesto,oliverviqarbh}.  From this one can construct 
a plausible answer to the problem of loss of information in black hole
evaporation\cite{abhaymartinbh}.  

 \blankline

  {\bf 22.} Suitable approximate calculations reproduce the Hawking
radiation. They further 
  predict a discrete fine structure in the
  Hawking spectrum\cite{kirill-radiate,fineBH}. At the same time,
  the spectrum fills in and becomes continuous in the limit of
  infinite black hole mass. 
  This fine structure stands as another definitive
  physical prediction of the theory.  

   \blankline
  
  Thus, to summarize, loop quantum gravity leads to a
  detailed microscopic picture of the quantum geometry of a black hole
  or cosmological horizon\cite{isolated}. This picture reproduces
  completely and  explains the results
  on the
  thermodynamic and
  quantum properties of horizons from the work of
  Bekenstein\cite{bek1}, Hawking\cite{hawking} and Unruh\cite{unruh}.
  This picture is completely general and applies to all black hole
  and cosmological horizons.

  \subsection{Results on the low energy behavior}
  
   {\bf 23.} A large class of putative ground states can be constructed in ${\cal H}^{kin}$
   which have course grained
  descriptions which reproduce the geometry of flat
  space, or any slowly varying metric\cite{weaves}-\cite{coherent}.  
  Linearizing the quantum
  theory around these states yields linearized quantum gravity, for
  gravitons with wavelength long compared to the Planck
  length\cite{graviton-weave}.  
  
   \blankline
  
  {\bf 24.} Excitations of matter fields on these states reproduce a cut off 
  version of the matter quantum field theories, but with a physical, 
  Planck scale cutoff. As a result of the discreteness of area and 
  volume, the ultraviolet divergences of ordinary $QFT$ are not 
  present\cite{thomas-thesis}. 
  
   \blankline
  
  {\bf 25.} It is understood rigorously how to
  construct coherent states in ${\cal H}^{kin}$ which are peaked around classical
  configurations\cite{coherent}.

   \blankline
  
  {\bf 26.} Formulations of the  renormalization group for 
    spin foam models have been given in \cite{f-RG,f-FRG}.  
    As a byproduct of this work it is shown that while the Wilsonian
    renormalization group is not a group, it does have a natural
    algebraic setting, as  a Hopf algebra.

   \blankline
  
  {\bf 27.} For the case of non-vanishing cosmological constant, of
  either sign, there is an exact physical state, called the Kodama 
  state, which is
  an exact solution to all of the quantum constraint equations, whose
  semiclassical limit exists\cite{kodama}.\footnote{A related state in the context
of Yang-Mills theory was previously discussed by Jackiw\cite{jackiw-kodama}.  Some
properties of Jackiw's state were discussed by Witten\cite{witten-kodama}.  More about this
in a FAQ.}
  That limit describes deSitter or
  anti-deSitter spacetime.  Solutions obtained by perturbing around
  this state, in both gravitational\cite{positive}
  and matter fields\cite{chopinlee}, reproduce, at
  long wavelength, quantum field theory in curved spacetime and the
  quantum theory of long wave length, free gravitational waves on
  deSitter or anti-deSitter spacetime\footnote{For a possible
  $\Lambda=0$ analogue of the Kodama state, see \cite{mikovic}.}.

 It is not yet known whether or not the Kodama state can be understood in a
rigorous setting, as the state is not measurable in the Ashtekar-Lewandowski
measure which is an essential element  of the rigorous approach to the canonical
quantization.  At the
same tine, the integrals involved are understood rigorously in terms of conformal
field theory.  It is thus not known if the state
is normalizable 
  in the physical inner product of the exact theory. One can ask 
  whether the projection into the Hilbert space of linearized 
  gravitons on deSitter spacetime gotten by truncating the Kodama 
  state to quadratic order is normalizable or not. The answer is it 
  is not 
  in the Lorentzian case, and it is delta functional normalizable in 
  the Euclidean case\cite{laurentlee}. 
  
   \blankline

  {\bf 28.} The inverse cosmological constant turns out to be quantized
  in integral units, so that 
  \f
  k= 6 \pi / G \Lambda
  \label{level}
  \ff
  is an
  integer\cite{linking}.  This is related to a basic result, which is 
  that in $3+1$ dimensions, a non-zero cosmological constant implies 
  a quantum deformation of the gauge group whose representations 
  label the edges of the spin networks, where the level which
  parameterizes the quantum deformation is given for the Euclidean 
  case by eq.
  (\ref{level})\footnote{and for the Lorentzian case by $\imath k$.}. 
  
   \blankline

  {\bf 29.} The thermal nature of quantum field theory in a deSitter
  spacetime is explained in terms of a periodicity in the
  configuration space of the exact quantum theory of general
  relativity\cite{chopinlee,positive}.
  
   \blankline

    {\bf 30.} In both flat space and around deSitter spacetime, one may
  extend the calculations that reproduce quantum theory for
  long wavelength gravitons and matter fields to higher energies.
 These calculations reveal the
  presence of corrections to the energy-momentum relations of the
  form of 
  \f
E^{2} = p^{2}+ M^{2} + \alpha l_{Pl}E^{3}+ \beta l_{Pl}^{2}E^{4}
+ ...
\label{modified}
\ff
Given a candidate for the ground state, the parameters
 $\alpha$ and $\beta$ are computable constants, that depend on
  the ground state wavefunctional\cite{GP,AMU,positive}.
Thus, given a choice for the ground state the theory yields predictions
for modifications of the energy momentum relations.   

 As discussed above, in the $2+1$ dimensional theory, modifications of the energy
momentum relations (\ref{modified})  are present.  They are 
understood as indicating, not a breaking
of Poincare invariance, but a deformation of it\cite{dsr2+1}.   It is 
conjectured\cite{positive,GAL},
but not shown, that the same will be true of quantum gravity in $3+1$ dimensions.   This will be
discussed further in the next section.  
  
  To summarize, the situation with regard to the low energy limit is 
  very much like that of condensed matter systems. It is possible to 
invent and study candidates for the ground state, which have reasonable physical 
  properties and reproduce the geometry of flat or deSitter spacetime. 
  By studying excitations of these states one reproduces conventional 
  quantum field theory, as well corrections to it which may be compared 
  with experiment.  
  
 There is not so far a first principles calculation that 
  provides an exact or unique form of the ground state wavefunction.  This is 
  also the case in most condensed matter physics examples. It remains an open question whether one can derive first principles or model independent physical predictions for the energy momentum relations  (\ref{modified}) from the theory. 
This will be discussed further below.  
  
\subsection{Results concerning cosmology}
 
  {\bf 31.} An approach to quantum cosmology has been developed from applying
the methods of loop quantum gravity to the spatially homogeneous case\cite{LQC}. 
This is called {\it loop quantum cosmology}, (LQC).   
There is a physical (but so far not rigorous) argument that the states studied
in this reduced approach correspond to spatially homogeneous states in the Hilbert
spaces,${\cal H}^{diff}$ of the full theory.  The space of states of the homogeneous
theory is motivated by the fact that normalizabie states of the full theory, based on Wilson loops, reduce
to normalizable states of the reduced theory. 

This means that the Hilbert space of LQC is not (for the simplest models, that
just depend on the scale factor, $a$) ${\cal L}_2 (R^+)$ as
is the case in most semiclassical approaches to quantum cosmology.
Instead, it is a {\it Bohr} quantization, in which $A$ is not a well
defined operator, but $e^A$ is.  As a result, the results differ
from those of older approaches to quantum cosmological models.  

\blankline

  {\bf 32.} The evolution of states of this reduced theory has been studied in
detail and it has been found generically that the usual FRW
  cosmology is reproduced when the universe is very large in Planck
  units. At the same time, the cosmological singularities are removed,
  and replaced by bounces where the universe re-expands (or
  pre-contracts).  

The replacement of cosmological singularities by bounces has  been confirmed  in several different models of this type, that make different assumptions as
to symmetries and dynamics, and hence appears 
robust\cite{GP-discrete,oliverviqarbh}.  

It is not yet known rigorously whether the singularities are eliminated in the full theory. But it has been shown that some of the properties of the density
operator in the LQC model are carried over into the full theory, including the
fact that the density operator is finite on states of the theory that have zero 
volume\cite{johannes-thomas-singular}

   \blankline

  {\bf 33.} When couplings to a scalar field are included,
  there is a natural mechanism which generates Planck scale inflation
  as well as a graceful exit from it\cite{LQC}.  Using this formulation of 
quantum cosmology it has  recently it has been 
  argued that loop quantum gravity effects lead naturally to a version 
  of chaotic inflation which may also explain the lack of power on 
  long wave lengths in the $CMB$ spectrum\cite{lqc-inflation}.
  
   \blankline

{\bf 34}  The transplanckian corrections to the spectrum of
fluctuations have been computed in one such model\cite{stefanoliver,hossain}.
At least in this one model, order $l_{Pl}H$ corrections vanish, while   
there are explicitly computer corrections at  order $(l_{Pl}H)^2$, where
$H$ is the Hubble constant during inflation. 

   \blankline

  {\bf 35.} Another approach to 
  inflation within loop quantum gravity is given in \cite{JSL}.  
  Exact homogeneous quantum states can be constructed for general 
  relativity coupled to a scalar field with an arbitrary potential.   This may be used to
construct exact quantum states corresponding to inflating universes, in the homogeneous
approximation. 
  
   \blankline

  {\bf 36.} It has  been shown that loop quantum gravity effects 
  eliminate the chaotic behavior of Bianchi IX models near 
  singularities\cite{martin-bianchi}.  

   \blankline

{\bf 37}  A mechanism analogous to the Pecci-Quinn mechanism has been proposed for relaxation of the cosmological constant\cite{stephon-cc}.  

\blankline
  
  \subsection{Results concerning supergravity and  other dimensions}
  
  {\bf 38.} Many of these results extend to quantum supergravity for $N=1$
  and several have been studied also for $N=2$ \cite{super,superyi,yime-holo}.
  
   \blankline

  {\bf 39.}  The same methods can
  also be used to solve quantum gravity in $2+1$ 
dimensions\cite{2+1}  and in some  $1+1$ dimensional 
reductions of the theory\cite{1+1}.
  They also work to
  solve a large class of topological field theories\cite{tft,BF},
  giving results
  equivalent to those achieved by other methods. Further, loop
  methods applied to lattice gauge theories yields results equivalent
  to those achieved by other methods\cite{latticeloop}. 
  
   \blankline
  
  {\bf 40.}  Spin foam models are known in closed form for quantum general 
  relativity in an arbitrary dimension $d>4$ \cite{higher}.

{\bf 41}  There is a class of diffeomorphism invariant theories
in $6$ and $7$ dimensions whose degrees of freedom are just 
forms\cite{hitchin}.  These have been conjectured to describe a diffeomorphism
invariant theory of interest to string theorists called {\it Topological
$\cal M$ theory\cite{topM}.}  This has been quantized using 
LQG methods\cite{me-topM}.

  \subsection{Other extensions of the theory}
  
  {\bf 42.} The mathematical language of spin networks and spin foams 
  can be used to construct a very large class of background 
  independent quantum theories of gravity. These may be called 
  generalized loop quantum gravity models. The states of 
  such a quantum theory of gravity are given
  by abstract spin networks\footnote{An abstract spin
  network is a combinatorial graph whose edges are labeled by representations
  of $\mathcal A$ and whose nodes are represented by invariants of
  $\mathcal A$. associated with the representation theory of a
  given Hopf algebra or superalgebra, $\mathcal A$}.  
  
  The graphs on which the spin networks are based are defined combinatorially,
  so that the
  need to specify
  the topology and dimension of the spatial manifold is
  eliminated\cite{F-foam,tubes}.
  In such a theory the dimension and topology are dynamical,
  and different states may exist whose coarse grained descriptions
  reveal manifolds of different dimensions and topology.

  The histories of the
  theory are given by spin foams labeled by the same representations.
  The dynamics of the theory is specified by evolution
  amplitudes assigned to the nodes of the spin foams (or equivalently
  to local moves by which the spin networks evolve).

  Such generalized loop quantum gravity models have been proposed to 
  serve as background independent formulations of string and $\mathcal M$ 
  theory\cite{stringnets,tubes}.  They also offer new possibilities 
  for the unification of physics, because the topology of spacetime is not
  assumed {\it a priori}, and hence must be emerge dynamically in the low 
  energy limit. There is then the possibility of new physics coming 
  from obstructions to recovering trivial topology at low energies. 
  It has been proposed that these may account for the existence of 
  matter degrees of freedom\cite{louis-matter,unify}, and perhaps even quantum 
  theory itself\cite{hidden-graphs}.

 \subsection{Comments on the results}

  On the basis of these results, it can be claimed that loop quantum
  gravity is both a consistent quantization of general relativity
and a physically plausible candidate for the quantum theory of gravity.

  The failure of quantum general relativity in perturbation theory is
  explained by the fact that there are, in this quantization of
  general relativity, no degrees of freedom that correspond to
  gravitons or other perturbative quanta with wavelength shorter than
  the Planck scale.  The ultraviolet divergences are eliminated
  because a correct quantization, that exactly realizes spatial
  diffeomorphism invariance, turns out to impose an ultraviolet cutoff
  on the physical spectrum of the theory.  The assumptions,  made by other approaches, 
   that spacetime is smooth and lorentz invariant 
  at arbitrarily short
  scales, are not used in the quantization procedure, and 
  in fact turns out to
  be contradicted by the results.

\section{The near term experimental situation}

The most important development of the last few years in quantum gravity
is the realization that it is now possible to probe Planck scale physics
experimentally\cite{GAC1}-\cite{wave}.  These experiments look for modifications in the
energy momentum relations of the form of eq. (\ref{modified}). 

However it is crucial to mention that to measure $\alpha$ and $\beta$ of eq. (\ref{modified}) 
one has to specify how lorentz invariance is treated in the theory. There
are two very different possibilities which must be distinguished.

\begin{itemize}
    
\item{}{\bf Scenario A)} The relativity of inertial frames is broken and 
there exists a 
preferred frame.
In this case the analysis has to be done in that preferred frame. 
The most likely
assumption is that the preferred frame coincides with the
rest frame of the cosmic microwave background.  In such theories energy and
momentum conservation are assumed to remain linear.

\item{}{\bf Scenario B)} The relativity of inertial frames is preserved, 
but the lorentz transformations
are realized non-linearly when acting on the energy and momentum eigenstates of
the theory. Such theories are called modified special relativity or 
doubly special
relativity\cite{gac-dsr,joaolee1,joaolee2,dsr2}. Examples are given by some forms of non-commutative
geometry, for example, $\kappa-$Minkowski spacetime\cite{kappa}.  
In all such theories energy and momentum conservation 
become non-linear which,
of course, effects the analysis of the experiments.

\end{itemize}
 
Among the experiments which either already give sufficient
sensitivity to measure $\alpha$ and $\beta$, or are expected to by 2010,  are,
\begin{enumerate}

        \item{}There are apparent violations of the GZK bound
        observed in ultra high energy cosmic rays (UHECR) detected
        by the AGASA experiment\cite{AGASA}. The experimental situation is
        inconsistent, but the new AUGER cosmic ray detector, which
        is now operational, is expected to resolve the situation
        over the next several years. If there is a violation of the
        GZK bound, a possible explanation is Planck scale physics
        coming from (\ref{modified}) \cite{AC-Piron}.  
	
	In Scenario A) violations of
	the GZK bounds can be explained by either $E^{3}$ or
	$E^{4}$ terms in the proton energy-momentum relation.
	However, in Scenario B) it is less natural
	to explain a
	violation of the GZK bounds by means of a Planck scale
	modification of the energy-momentum relations, but there
	are proposals for forms of such theories that do achieve this. 

       \item{}A similar anomaly is possibly indicated in Tev
       photons coming from blazers\cite{blazers}.  Similar remarks
       apply as to the explanatory power of Scenarios A) and B) in the 
       event that the anomaly exists. 

        \item{}A consequence of (\ref{modified}) can be an energy
        dependent speed of light. This effect can be looked for in
        timing data of gamma ray busts. Present data bounds
        $\alpha < \approx 10^4$ \cite{wavelets} 
	and data expected  from the GLAST
        experiment is expected to be sensitive to $\alpha$ of
        order one in 2006 \cite{glast}. Note that this applies to both
	Scenarios A) and B). 

        \item{}Present observations of synchrotron radiation in the 
	Crab nebula, together with reasonable astrophysical assumptions,
	put {\it very strong} (of order $10^{-9}$)
        bounds on $\alpha$ for photons and electrons, in the case
        of Scenario A only\cite{synch}.

        \item{}Present data from precision nuclear and atomic
        physics experiments puts very tight bounds on $\alpha$ for
        photons, electrons and hadrons, again in Scenario A),
        only\cite{atomic}.

        \item{}Present data from the absence of vacuum Cherenkov
        effects puts interesting bounds on $\alpha$ in the case of
        Scenario A) \cite{seth}.

        \item{}Observations of bifringence effects in polarized light from
        distant galaxies puts tight
        bounds on a possible helicity dependent $\alpha$  
        \cite{helicity}.

       \item{}Observations of phase coherence in stellar and 
	galactic interferometry
        is expected, given certain assumptions\footnote{See \cite{ng} 
        for discussion of them.}, 
	to put order one bounds on $\alpha$ in the 
	near future \cite{interference}.

        \item{}Certain hypotheses about the Planck scale lead to
        the prediction of noise in gravitational wave detectors
        that may be observable at LIGO and VIRGO\cite{wave}.

       \item {}Under some cosmological scenarios, modifications of
        the form of (\ref{modified}) lead to distortions of the
        CMB spectrum that may be observable in near future
        observations\cite{CMB-distort}.

\item{} There is a possibility that observations of very high energy 
neutrinos from cosmological distances, in experiments such as ICEBERG may make it possible to test for both violations of special relativity and loss of quantum
coherence\cite{joy-neutrinos}.

\end{enumerate}

We may summarize this situation by saying that a theory of quantum
gravity that leads to Scenario A) and predicts an energy momentum
relation (\ref{modified}) with $\alpha$ order unity is plausibly
already ruled out. This is shocking, as it was commonly said just a few
years ago that it would be impossible to test any physical
hypotheses concerning the Planck scale.

We can also mention three other kinds of experiments 
that by 2010 will have relevance for the
problem of quantum gravity

\begin{enumerate}

\item{}Evidence for or against supersymmetry may be detected at the Tev scale
at the LHC.

\item{}The equation of state of the dark energy will be measured
in near future experiments. Some proposals for dark energy\cite{laura-dark} are based on
modifications of energy momentum relations of the form of (\ref{modified}).

\item{}There are observations that appear to indicate that 
the fine structure constant is time
dependent\cite{fine}. These will be confirmed or go away. If the claim is
substantiated this offers a big challenge to the effective field theory
understanding of low energy physics.

\end{enumerate}

The combination of all these experimental possibilities signals that the
long period when fundamental physics developed independently of
experiment is soon coming to a close.

\subsection{What does loop quantum gravity predict for the 
  experiments?}

It is first important to observe 
that there is no reason that  the low energy limit of quantum gravity must be
Poincare  invariant. Poincare invariance is not a symmetry of general
relativity, it is only a global symmetry of a particular
solution of the classical  theory.
Global symmetries including Poincare and Lorentz invariance 
are not symmetries of the fundamental
theory of gravity, neither classically nor quantum mechanically.  
Whether these symmetries are fully realized in the ground state
has to be determined by calculation.

Indeed, several recent calculations, done with
  different methods\cite{GP,AMU,positive},
  yield predictions for modifications to the energy
  momentum relations for elementary particles of the form of
  equation (\ref{modified}).
Predictions have been found for the leading coefficients $\alpha$,
which generally depend on spin and helicity\cite{GP}-\cite{positive}.

One issue here is that different
calculations make different assumptions about the ground state. In
some the ground state is not Lorentz invariant, hence there is no 
surprise if the perturbations around them have non-lorentz invariant
spectra. However, modified dispersion relations may also be seen 
by studying low energy excitations of a putative around state that does 
not single out a preferred 
frame\cite{positive}. The question is then: can we determine
the ground state precisely enough to discover whether the theory makes
unambiguous predictions for the parameters in the energy-momentum
relations (\ref{modified})?  Of particular importance is to determine
which scenario is realized, A) or B) ?. 
  
 There is a straightforward
 argument leading to an expectation that  Scenario B) is realized. This
  is that, even though there is no global lorentz invariance in
  classical general relativity, the existence of effects due
  to a preferred frame are ruled out by the condition of invariance
  under the action of the hamiltonian constraint. This is because
the hamiltonian constraint can generate
  changes in slicing that in any finite region are
  indistinguishable from lorentz boosts.  

  Now some of the key results of loop quantum gravity tell us that 
  the hamiltonian constraint can be defined and solved exactly, and
  that no anomalies are introduced into the constraint algebra by the
  quantization. This implies that any quantum
  state that is both an exact solution to the hamiltonian constraint
  and has a semiclassical limit will in that limit describe
  physics which is to leading order invariant under the action of
  the classical hamiltonian constraint. This implies the absence
  of a preferred frame of reference in the classical limit of an
  exact solution to the hamiltonian constraint.

The only exception to this could be if there were a vector or tensor field in the theory that could pick
up a vacuum expectation value in the ground state, essentially breaking Lorentz invariance
spontaneously.  The field in question must be able to get a vacuum expectation value without
breaking any gauge invariance. Thus, it 
cannot be the metric field itself, for if it had an expectation value it would
break also diffeomorphism invariance. This cannot happen as diffeomorphism invariance
is a gauge symmetry. Nor can it be an ordinary gauge field.  Thus,spontaneous breaking
of Lorentz invariance cannot be generic, it can only be due
to a special matter dynamics being imposed on the theory.  
  
  So this appears to rule out Scenario A), in all but a few artificial cases. 
  
This reasoning is supported by many results on $2+1$ gravity, 
which show in detail that scenario B) is realized in that case\cite{dsr2+1}. 

Although most results on $2+1$ gravity do not extend to the physical,
$3+1$ theory, there is an argument for scenario B that does appear to 
extend to $3+1$\cite{GAL}. This argument starts with the theory with non-zero 
cosmological constant. In $2+1$ gravity, with non-zero $\Lambda$, 
the symmetry group of boundary states 
is $SO_{q}(3,1)$, the quantum deformed deSitter group, with deformation
parameter given in the $2+1$ dimensional case by
\f
\ln  q \sim l_{P} \sqrt{\Lambda}
\label{2+1level}
\ff
When one takes the contraction in which the cosmological constant 
goes to zero one does not get, as usual the Poincare algebra. Instead,
because $q \rightarrow 1$ as $\Lambda \rightarrow 0$ one gets a 
deformation of the Poincare algebra parameterized by the Planck 
length $l_{P}$.  This is the $\kappa$-Poincare algebra, which is 
known to provided an example of Scenario B. The emergence of 
$\kappa$-Poincare symmetry in $2+1$ gravity is also confirmed in 
several detailed studies\cite{dsr2+1}.  

One can run the same argument in $3+1$ dimensions, but one must use 
the relation (\ref{level}) instead of (\ref{2+1level}).  Because the 
theory has local field degrees of freedom, one must also renormalize 
the energy and momentum generators by an appropriate power of
$l_{P}\sqrt{\Lambda}$, which is the ratio of the physical ultraviolet 
and infrared cutoffs. When this is done, one recovers again the 
$\kappa$-deformation of the Poincare algebra\cite{dsr2+1}. 

Remarkably, evidence for the same result is found in studies
of excitations of the Kodama state \cite{positive}.  However, this result has
been criticized because  it is not known whether or not the Kodama state
is a normalizable physical state. To counter this worry, the argument has
been recast to show that it arises from studying excitations of a general
semiclassical state\cite{semi-me}.

\section{Frequently asked questions}

There is a set of questions often asked by physicists on first encountering loop quantum
gravity.  For the convenience of the readers, some of these are collected here, with answers.

\begin{enumerate}

\item{}{\it How can there be a finite, well defined formulation of quantum general
relativity when that theory is not normalizable in perturbation theory?}  The reason is
that the standard perturbative approaches make two assumptions which are not made
in the exact approach followed in LQG. i)  Spacetime is smooth down to arbitrarily short distances, so there are physical degrees of freedom which propagate for arbitrarily high frequency and short wavelength. ii)  The standard lorentz transformations correctly apply to these modes, no matter how high the frequency. Neither assumption  could be made in a background independent approach. Indeed, the results of LQG falsify the first assumption and make testable the second.  Physically speaking, there simply are no weakly coupled excitations of 
gravitational or matter fields with wavelength shorter than $l_{Planck}$.  

\item{}{\it Can the same methods be applied to ordinary quantum field theories?}
In fact some of the results of the loop representation were found first in 
lattice gauge theories, such as 
the fact that the spin networks give an orthonormal
basis\cite{lattice-old}.  An approach to solving $QCD$ numerically, with
fermions, has been developed using the methods of loop quantum gravity\cite{latticeloop}.
However, as mentioned above, spin network states in the continuum give
a non-separable Hilbert space, unless one mods out the state space by the
action of the diffeomorphisms. Thus, they can be used in two contexts,
either in the presence of a lattice regularization or in a gravitational theory
in which the physical states are diffeomorphism invariant. 

\item{}{\it Can LQG methods be applied to string theory?} Background
independent methods of quantization, analogous to LQG have been
applied to the free string in \cite{artem-string,thomas-string,HP-string}.  These
give theories which are well defined but which are unitarily inequivalent
to the usual quantization of the free string. The reason is that the usual
quantization uses a representation of $Diff(S^1 )$ which is anomalous,
whereas the LQG-type quantizations use non-anomalous representations.
The former approach uses structure special to $1+1$ theories which is that
there is a generator of $Diff(S^1)$, called $L_0$, which in a certain
gauge can be interpreted as the hamiltonian of the free string. It
makes sense, in the quantization of a gauge fixed quantization, to choose
a representation in which $L_0$ is positive and hermitian.  The background
independent quantizations on the other hand must treat all the generators
of $Diff(S^1 )$ as constraints.  Thus, the two methods are different
and yield different theories. This does not mean one or the other is right,
but it does illuminate an important difference between background
independent and background dependent approaches to quantization. 

\item{}{\it LQG as well as finite dimensional models based on it studied
in Loop Quantum Cosmology depend on a non-standard representation
of the quantum theory in which the canonical commutation relations
are not represented. Instead, one quantizes commutation relations
involving  exponentials of the basic variables.  When this approach is applied to  ordinary systems such as the harmonic oscillator it gives a theory that is
unitarily inequivalent to the standard quantum theory.  Why then should be
trust it when applied to gravity?}   For field theories, the standard quantization,
which leads to Fock space, depends on a fixed background metric. So it is
simply unavailable for the case of a background independent approach to
quantum gravity.  A new approach is needed, which must be based on the
quantization of a non-canonical algebra.  This is what is done in loop quantum
gravity.  The fact that ordinary QFT is reproduced in suitable limits, as discussed
in results above shows that the standard results of quantum theory are 
reproduced when suitable approximations are made, leading to the emergence
of physics on a background spacetime.

\item{}{\it  Why isn't supersymmetry necessary, when string theory says it is for
perturbative finiteness of quantum gravity?  }   The answer is straightforward. Because  there are no infinite sums over arbitrarily high energy modes,  there is no
need for supersymmetry to cancel them, by balancing the contributions of 
fermions and bosons. 
There are, so far,  
no results that indicate that matter context effects the consistency or finiteness
of the quantum theory of gravity.  Loop quantum gravity appears 
equally finite and consistent coupled to any matter fields. At the same time there 
is no problem extending all the main results to $N=1$ supergravity\cite{yime-holo}-\cite{11d}
There are
interesting problems extending loop quantum gravity to $N=2$ and higher, which
have yet to be tackled. 

\item{\it Do you have anything to say about the problem of unification?} 
There are arguments that matter is necessarily included in loop quantum
gravity, as a result of mismatches between microsoiopic and macroscopic
notions of locality\cite{unify} and topololgy\cite{louis-matter}.

\item{\it Do you have anything to say about the problem of the cosmological
constant?}YES.   There is a proposal for a mechanism to dynamically tune the
cosmological constant to zero, analogous to the Pecci-Quinn 
mechanism\cite{stephon-cc}. 
It can also be mentioned that, unlike approaches that depend on supersymmetry, 
there is no special issue incorporating a finite cosmological constant of either sign
into the theory.  There is also an argument that the problem of
the cosmological constant is a pseudo-problem brought on by background
dependent approaches, brought on by an unphysical split of the fundamental degrees of freedom into two sets, which represent the background and
excitations of the background\cite{olaf-cc}. 

\item{}{\it What about the problem of the existence of the low energy limit?}
As described above, there is no problem studying the low energy limit using
techniques analogous to those of conventional theories. One can easily write
down candidates for the ground state, and show that they have several properties required
for the ground state.  For example, one can study small excitations of these states and
recover ordinary quantum field theory on background manifolds, 
as well as the linearized graviton states, 
at least in the limit
of long wavelength.   The difficulties that remain with the low energy limit  have mostly
to do with finding criteria to distinguish the physically correct ground state from other 
candidates. 

\item{}{\it   Can you calculate graviton-graviton scattering? } In principle yes.
One could easily study interactions among the excitations of candidate ground states 
such as those in \cite{weaves,newweaves,morevacuum,coherent,positive}.  
 The result must be finite as the discrete of area and volume
impose  an ultraviolet cutoff on the spectra of excitations of any candidate ground
state.  Presumably the results will depend on the properties of the
ground state, and a crucial issue will be the fate of Lorentz invariance.  We would not be
surprised to see that the scattering of excitations around ground states of the form
of \cite{weaves} that depend on preferred frames break Lorentz invariance explicitly. 
Still, it is embarrassing that this has not been done for some examples, as it would
be illuminating to see exactly how perturbative finiteness arises. 

\item{}{\it In string theory we take {\it background independence} to mean independent of the 
particular vacuum perturbative string theory is defined with respect to. Is this the same 
meaning you use in LQG?}  No, the key point in the definition given in section 2 above is that the
quantum theory is completely defined without reference to any fixed background fields.
This is different from a formulation that requires a background field, even if there may be
many different background fields that will serve\footnote{For more on the
meaning of background independence, see \cite{background}.}.

\item{}{\it What about the dimension and topology of the manifold. Shouldn't that be
determined dynamically as well?}  The general philosophical principle that historically
motivates background independence is the idea that space and time should be purely
relational quantities. This does imply that one who believes that principle would prefer
that topology and dimension are emergent and dynamically determined quantities.
However, this is not the case in general relativity and other classical diffeomorphism
invariant field theories. While one can study general relativity for different topologies,
dimensions and choices of boundary conditions, each set of these choices defines a
phase space. The different phase spaces corresponding to the different choices of dimension,
topology and differential structure are disconnected from each other.  To construct a phase
space the topology must be chosen to be of the form of $\Sigma \times R$, with 
the spatial manifold $\Sigma$ fixed, hence topology change cannot be described in
a canonical formalism. In a Hamiltonian quantization there is then a different Hilbert 
space for each choice of topology, dimension and boundary conditions, so these are fixed
as well within each quantum theory\footnote{This issue is discussed in 
more detail in \cite{background}.}.

This circumstance does not prevent people from taking the structures from loop quantum
gravity and proposing new theories in which the states and histories are defined in
terms only of combinatorial graphs and representation theory, so that the dimension
and topology may emerge as aspects of the low energy behavior.  There might
then be phase transitions in which topology and dimension change., Such a theory could
not be strictly speaking the ``quantization of a classical field theory", but some believe
that makes it more rather than less likely to describe nature. These theories are
mentioned in 4.8 above.

\item{}{\it    Could string theory and loop quantum gravity be different sides of the same
theory?}   It is natural 
to make this suggestion as both string theory and loop quantum gravity realize the principle 
of duality-the former in a background dependent context and the latter in a background
independent context.  This makes it natural to hypothesize that excitations around 
background loop quantum gravity states whose semiclassical approximations look
like smooth spacetimes should behave like  perturbative strings, at least in a regime
below the Planck scale.  This possibility has been explored
in a number of papers\cite{Mlee}.  The results look intriguing but there is more to do. 
Given that string/$\cal M$ theory needs a background independent formulation, and given
that the most developed approach to quantum gravity at the background independent
level is loop quantum gravity, it is natural to look for a version of loop quantum gravity
that could serve as the background independent form of string theory.  For example,
one can study spin foam models based on the conjectured symmetries of $\cal M$
theory such as $Osp(1|32)$.  One approach to unifying string theory and loop
quantum gravity by subsuming both in a certain extension of a matrix formulation
of Chern-Simons theory is described in \cite{Mlee}.  

\item{}{\it I've heard that the Kodama state has been shown to be unphysical.  Is this true?}
Some criticisms have been made concerning a related state in Yang-Mills theory, discussed
by Jackiw in \cite{jackiw-kodama} and Wittin in \cite{witten-kodama}.  It is likely that this state is
unphysical in the Hilbert space of Yang-Mills theory for the reasons discussed by those authors. 
However the analogy to quantum general relativity is not very helpful here, for reasons discussed
in \cite{laurentlee}.    For one thing, the Yang-Mills version does not play the role of a semiclassical
state to describe excitations around the classical ground state, as is the case in gravity.  
The precise situation is as was described in result {\bf 27} above.  The Kodama state is an
exact solution to the constraints of quantum gravity and is also a semiclassical state corresponding
to an expansion around de Sitter (or AdS) spacetime\cite{kodama}.  It is not known whether
or not the Kodama state is normalizable in the physical inner product.  A naive truncation 
results in a linearized state that is not normalizable in the lorentzian case and delta-functional
normalizable in the Euclidean case\cite{laurentlee}.  Semiclassical expansions around the Kodama state
lead to a recovery of QFT on deSitter (or AdS) spacetimes\cite{chopinlee,positive}.

\item{}{\it Does loop quantum gravity have anything to say about the holographic principle?}
Quite a bit. It should be mentioned that Crane, in \cite{louis-holo} made a conjecture for 
a background independent formulation of the holographic principle, a bit before the papers of 
't Hooft\cite{thooft-holo} and  Susskind\cite{lenny-holo}.  Crane's
formulation inspired the study of Chern-Simons boundary terms\cite{linking}, which then led
to the studies of black hole entropy.  In fact, the $N$-bound discussed in \cite{N-Banks} can
be explained in terms of loop quantum gravity, as discussed in \cite{positive}.  A background
independent formulation of the holographic principle, suitable for theories in which the spacetime
geometry is fully quantized was proposed in \cite{weakholo} and compared to other formulations 
in \cite{weakstrong}.

\item{}{ \it Isn't it true that the Ashtekar-Sen connection lives in a complex extension of $SU(2)$?
Doesn't this mean that the inner product cannot be defined so that one has just traded one
set of problems for another?}  This is wrong on two counts. First, most results listed above
are based on the use of the Barbero connection, which is valued in the real $SU(2)$.  The
inner products on ${\cal H}^{kin}$ and ${\cal H}^{diffeo}$ are then rigorously defined
and a full normalizable basis is known explicitly in each case.   Thiemann's form of the 
Hamiltonian constraint applies to this case (for both the Lorentzian and Euclidean signature.) The Hamiltonian
constraint is finite and well defined, acting on ${\cal H}^{diffeo}$ and an infinite number
of exact solutions are known.  

There continues to be some interest in the Ashtekar-Sen connection.  Not as many rigorous
results have been achieved for it, but the problem is not that of trying to make a Schroedinger-like
representation on a space with a non-compact gauge field.  For example., there is no problem at all working out the details of the linearized theory of spin-two fields in terms of the
linearization of the Ashtekar connection\cite{gravitons,laurentlee}.  
The reason is that the Ashtekar-Sen connection is
a non-linear version of a Bargmann coordinate, $z=q+ip$ for an ordinary quantum system.
The states are then holomorphic functionals of $A_a^i$.   The key open problems concern the
normalizability of such states under the physical inner product. So far it has been easier
to address such problems in the context of the Barbero connection, but the problem 
remains open.

\item{}{\it I've heard there is some problem with the consistency of the Hamiltonian
constraints\cite{outsideview}. What was it, and has it been resolved?}   There is no problem with the
consistency of the constraints.  To explain why is a bit technical, but given the
importance of the issue I will go into it.  The issue has to do with how the classical
algebra of the constraints of general relativity is realized quantum mechanically\cite{problem1,problem2}. 
Classically, the Poisson brackets of two Hamiltonian constraints, $H(x)$, of the form of 
(\ref{hamiltonianconstraint}) have the
form
\f
\{ H(x) , H(y) \} = \partial_a \delta^3 (x,y) qq^{ab} D_b
\label{class-algebra}
\ff
where $D_b$ is the diffeomorphism constraint, (\ref{diffconstraint}).  The problem is
that  this is not a Lie algebra because the structure constant of the algebra 
includes  $q^{ab}$, a dynamical variable\footnote{This is sometimes referred
to as a business class, rather than a first class algebra.}.
The problem is then  what happens to the algebra in the quantum theory?  There
are several aspects to this problem.  First, the operators are all regularized in the
quantum theory, so $H(x)$ is realized by a one parameter family
$\hat{H}_\epsilon$.   The condition that a state be physical is that
\f
\lim_{\epsilon \rightarrow 0} \left ( \hat{H}_\epsilon (x) \Psi \right ) =0.  
\label{regulated}
\ff
This limit must be taken in a suitable norm, which turns out to be the inner 
product of the gauge but not diffeomorphism invariant states (because the
regulated $\hat{H}_\epsilon$ is not diffeomorphism covariant for finite $\epsilon$, and
only becomes so in the limit $\epsilon \rightarrow 0$.)

For the theory to exist, there must be an infinite number of simultaneous solutions
to (\ref{regulated}) which also live in the space $H^{diffeo}$ of
diffeomorphism invariant states. In fact, there are, given a completely
explicit construction of $\hat{H}_\epsilon$, infinite numbers of such
solutions have been constructed explicitly. 

The question is then what is the right hand side of,
\f
[\hat{H}_\epsilon (x) , \hat{H}_{\epsilon^\prime} (y) ]   = ? 
\label{rhs}
\ff   
From the
existence of an infinite number of solutions to (\ref{regulated})  
no anomalous c-number term can appear on the right hand side.   So the
algebra is consistent.  

In general, what appears on the right hand side depends on the regularization
and ordering studied.  In the most studied approach, due to Thiemann, the
right hand side vanishes in the limit, when evaluated on diffeomorphism invariant
states.  This of course agrees with the expectation from (\ref{class-algebra}) and it
is enough to ensure the consistency of the theory. However, a simple, but quantum
mechanical form of (\ref{class-algebra}) is not easily obtained in the space
of kinematical states ${\cal H}^{kin}$. Whether or not this is a problem is not clear. 

 In other regularizations, such as that studied
originally in \cite{tedlee}, the right hand side appears formally in the order 
$\hat{D}_a \hat{q}^{ab}$.  This makes it possible to find solutions to
(\ref{regulated}) which are not diffeomorphism invariant. Instead they 
satisfy  $\hat{D}_a \hat{q}^{ab} \Psi =0 $.  This turns out to mean that the states need
only be invariant under diffeomorphisms tangent to the loops, which they are by
virtue of the reparameterization invariance of Wilson loops. 

What is important is that  there is no anomaly or inconsistency.
A few  practitioners would still prefer a regularization
and ordering such that the right hand side (\ref{rhs}) reproduced a term non-vanishing
in the limit with the ordering  $ \hat{q}^{ab} \hat{D}_a  $. This would of course imply
that only diffeomorphism invariant states could satisfy (\ref{regulated}), (when  integrated
over all smooth test functions.)  So far none is known. However, there is not to my knowledge
a compelling case for requiring this. What is necessary, is to insure that the theory
has propagation of physical degrees of freedom. One way to ask this is the following:
Are there solutions to the full set of constraints which describe gravitons propagating
on classical backgrounds?  If so the theory contains the quantization of the physical
degrees of freedom of general relativity.  Examples of such solutions are described
in \cite{positive}.  

Another response to this issue has been to suggest that there is no need for
a theory of quantum gravity to have a generator of infinitesimal time
evolution if it has generators of finite evolution.  Indeed, given that there is
on ${\cal H}^{kin}$
no generator of infinitesimal spatial diffeomorphisms, but only a generator
of finite diffeomorphisms, it would contradict relativity were there to be 
a well defined generator of infinitesimal  time translations.  This argument was one of the basic motivations for the invention of spin foam models and justifies the
inclusion into the spin foam models of moves which are not present in
Thiemann's constraint\cite{F-foam,RR-foam}.

\end{enumerate}

\section{Open questions in loop quantum gravity}

I close with a brief list of some problems of 
current interest.

\begin{itemize}
    
    \item{}There is an urgency to determine what 
    unambiguous predictions can be made for the results of upcoming 
    experiments on Planck scale phenomenology. 
    
    \item{}A similar situation 
    holds for soon to be measured features of the CMB spectrum, such 
    as polarization and the tensor modes. 
    
    \item{}Gravitons and other quanta are known to arise as weakly 
    coupled excitations of several different ansatzes for the ground 
    state. As there are no modes below the Planck length, the 
    scattering of these excitations must be computable and finite. 
    It would be good to calculate some examples of scattering and 
    confirm this expectation value in detail. The answer will of 
    course depend on the ansatz chosen for the ground state.
   
    \item{}In contexts with a boundary, where there is a classical 
    positive definite energy, is there an expression for the 
    expectation value of the energy which is positive definite, thus 
    extending to the quantum theory the positive energy theorem? 
    
    \item{}Were there such a positive definite expression for the 
    energy, one could determine which is the right ansatz for the 
    ground state by minimizing it, either exactly or approximately. 
    This would lead to unambiguous predictions for Planck scale 
    phenomenology. An alternative way to do this would be to establish 
    a general result about the symmetry of the ground state, i.e. is 
    it invariant under the ordinary or deformed Poincare algebra? This 
    would also lead to predictions for scattering of gravitons and 
    other quanta. 
    
\item{}In conventional QFT we have learned that good theories
are associated with critical phenomena in statistical system, connected
to the path integral by Euclidean continuation.   Given that euclidean
continuation is not necessarily available in quantum gravity, one can ask whether there is a characteristic critical phenomena associated with quantum gravitational
systems\cite{ml-soc}.

\item{} There are indications that LQG contains matter degrees of
freedom intrinsically, so they do not have to be put in, which
need more development\cite{unify,sundance,louis-matter}.

    \item{}Dreyer's work concerning the relationship between black 
    hole entropy and quasi normal modes\cite{olaf-bh} is reassuring, because it 
    shows that the same value of the Imirzi parameter (or finite 
    renormalization of Newton's constant) suffices to match both 
    properties of black holes. But it leads to a puzzle: does this 
    mean that loop quantum gravity with the parameter chosen 
    differently cannot have a good classical limit?  Is the correct 
    value an attractor or a fixed point?  
    
    \item{}Loop quantum gravity can be extended to incorporate 
    supersymmetry. But is there anything that supersymmetry does for quantum 
    gravity at a background independent level, that would lead to a 
    reason to prefer a supersymmetric theory?

    \item{}Loop quantum gravity has its simplest formulation in terms 
    of the chiral Ashtekar-Sen variables. The classical theory is 
    nevertheless chirally symmetric, but it was suggested a long time 
    ago by Soo\cite{chopin-thesis} that there might be violations of CP and chirally 
    asymmetric effects quantum mechanically, and evidence for such 
    effects was found in studies of one ansatz for the ground state\cite{GP}.
    It would be good to know if LQG makes unambiguous predictions 
    about chiral symmetry breaking and $CP$\footnote{Some recent
results about this are in \cite{parity}.}.
    
    \item{}There are some technical problems remaining in 
    establishing the detailed connection between the Hamiltonian and 
    path integral, or spin foam, formulations. There are similarly 
    technical issues related to the exact relationship between the 
    different approaches to spin foam models. 

\item{} It is known that dynamical triangulation models have no 
good continuum limit in the case of $4$, Euclidean dimensions\cite{dynamical}, but
there is recent evidence that a lorentzian variant has a good low energy
limit  in $3+1$ dimensions\cite{AL,AL3+1}.  One can conjecture that the
same is true for spin foam models, i.e. that there will be a good
low energy limit only in the causal or lorentzian case. To show this
one might start by adding labels to CDT models, so as to turn them
into spin foam models.  It will be also necessary to 
extend the CDT models
so that they do not depend on a rigid time slicing. This seems to be
possible, at least in $1+1$\cite{fl-cdt} and $2+1$ dimensions\cite{tomasz-cdt}.
A useful step in this direction is also to re-express the CDT models as
spin systems\cite{f-m-CDT}

    \item{}While the physical inner product\footnote{By which is meant 
    the inner product on the subset of states that satisfy all the 
    constraints. The inner product on spatially diffeomorphism 
    invariant states is known in closed form.} is known formally to be expressed 
    as a path integral in the spin foam language, it would be good to 
    have it in a more explicit form.  This would allow us to check 
    whether physical states such as the Kodama state are normalizable 
    in the exact physical inner product. 
    
    \item{}More work needs to be done on the renormalization group
for spin foam models\cite{f-RG,f-FRG}. 
    
    \item{}There remain foundational issues having to do with the 
    wavefunction of the universe. The fact that we can do quantum 
    cosmology in detail, with real local degrees of freedom, rather 
    than just study models, makes the resolution of these issues more urgent. 
   Loop quantum gravity has stimulated new ideas in this direction, such as 
quantum causal histories\cite{Fotini-Wheeler,algebraic-fotini} and a proposal 
that spin networks might underlie a theory of hidden variables\cite{hidden-graphs}.
    
\end{itemize}    

\section{Conclusions}

The spirit of this review has been to lay out the results so that the reader can draw their own
conclusions.  In case it is interesting, the author would summarize the present state as
knowledge as follows.   

The open problems are non-trivial and the results so far by no means are sufficient
to show that loop quantum gravity is the right theory of quantum gravity. But the results are
sufficient to say that diffeomorphism invariant quantum gauge field theories
exist, rigorously, in both canonical and path integral form. 
The resulting theories are background independent, as any plausible quantum theory of
spacetime must be.  A powerful toolkit
of calculational techniques has been developed and many non trivial, and in some
cases, surprising results have been found. Among these theories are the
quantization of general relativity  coupled to the standard matter fields and supergravity. 

These theories are finite quantum field theories and the remaining open problems
concern, not their existence, but whether they reproduce general relativity correctly in the low
energy limit and make correct predictions for experiments that probe beyond that
limit. This is to say that loop quantum gravity is now part of ordinary physics in that future
progress will result from refinement of calculational techniques and models and
comparisons of the results gotten with experiment. It is possible, but by no means certain,
that the result of this process will be an accumulation of evidence that loop quantum
gravity correctly describes the physics of the Planck scale.  

While other approaches to quantum gravity also have gotten impressive results, 
there is not another approach to quantum gravity about which all of the foregoing can be
said\footnote{The best developed alternative, string theory, has also impressive
results, but  those concerning the consistent unification of general relativity with quantum
theory are limited so far to second order (genus two) perturbation theory around 
fixed backgrounds\cite{howfar}.}, or about which there is such a long list of robust results
concerning Planck scale physics. 

There is much still to do, and the sentiment expressed by the title is genuine.  
Loop quantum gravity may still not be right. But it can, perhaps, be said objectively that it is at
least, among the approaches so far known,  the most reasonable point of departure for
the discovery of the right theory of quantum gravity.

\section*{Acknowledgments}

I  must thank my collaborators over the
years who have taught me most of what I know about quantum gravity,
Stephon Alexander, Matthias Arnsdorf, Abhay Ashtekar,
Roumen Borissov, Louis Crane,  John Dell,
Ted Jacobson, Laurent Freidel, Jurek Kowalski-Glikman, Yi Ling, Seth Major,
Joao Magueijo, Justin Malecki, Fotini Markopoulou, Carlo Rovelli, Chopin Soo and Artem Starodubtsev.  Conversations with many other people during the preparation of this review 
have been very helpful, especially Jan Ambjorn, Olaf Dreyer, Renate Loll and Thomas
Thiemann.  
I am grateful as well to
the NSF and the Phillips Foundation for their very generous support which
made my own work possible.  
I am thankful especially to my colleagues at Perimeter Institute for many critical discussions.  Finally, I am grateful to John Schwarz for pointing out the need for this review, and to Thomas Thiemann for a careful and critical reading of the manuscript.

\end{document}